\pdfoutput=1
\documentclass[acmsmall,screen,nonacm]{acmart}
\AtBeginDocument{}
\setcopyright{acmlicensed}
\copyrightyear{2027}
\acmYear{2027}
\acmDOI{XXXXXXX.XXXXXXX}
\acmJournal{PACMPL}
\acmVolume{11}
\acmNumber{OOPSLA1}
\acmArticle{1}
\acmMonth{4}
\usepackage{listings}
\usepackage{algorithm}
\usepackage{algpseudocode}
\definecolor{codecomment}{gray}{0.45}
\definecolor{codekw}{RGB}{0,90,160}
\definecolor{trapred}{RGB}{200,20,20}
\definecolor{checkpink}{RGB}{252,222,232}
\usepackage{tcolorbox}
\newlength{\chkcw}\newlength{\chkpad}
\AtBeginDocument{\settowidth{\chkcw}{\footnotesize\ttfamily 0}\setlength{\chkpad}{2pt}}
\newenvironment{chkbox}[2]{\par\vspace{1pt}\raggedright\noindent
  \hspace*{\dimexpr 0.5em+#1\chkcw-\chkpad\relax}%
  \tcolorbox[nobeforeafter, colback=checkpink, colframe=checkpink,
    boxrule=0pt, arc=3pt, boxsep=0pt, left=\chkpad, right=\chkpad,
    top=1.5pt, bottom=1.5pt, width=\dimexpr #2\chkcw+2\chkpad\relax]}%
  {\endtcolorbox\par\vspace{1pt}}
\lstdefinestyle{odessy}{
  basicstyle=\footnotesize\ttfamily,
  escapeinside={(*}{*)},
  keywordstyle=\color{codekw}\bfseries,
  commentstyle=\color{codecomment}\itshape,
  columns=fullflexible,
  keepspaces=true,
  breaklines=false,
  xleftmargin=0.5em,
  aboveskip=0pt,
  belowskip=0pt
}
\newcommand{\odessy}{ODeSSy}
\newcommand{\tool}{\odessy}
\newcommand{\pct}[1]{#1\%}
\newcommand{\den}[1]{[\![#1]\!]}

\begin{document}
\title[Super-Analysis for Compiler Optimization]{Super-Analysis for Aggressive yet Scalable Compiler Optimization}
\author{Amirali Ebrahimzadeh}
\affiliation{\institution{University of Michigan}\city{Ann Arbor}\state{MI}\country{USA}}
\email{amiralie@umich.edu}
\author{Amir Shaikhha}
\affiliation{\institution{TU Darmstadt}\city{Darmstadt}\country{Germany}}
\email{amir.shaikhha@tu-darmstadt.de}
\renewcommand{\shortauthors}{Ebrahimzadeh and Shaikhha}
\begin{abstract}
Aggressive compiler optimization today occupies two distant extremes.
Conventional optimizers run in milliseconds but reason through ad-hoc
heuristics that miss opportunities; superoptimizers synthesize faster code
but take hours, confining them to small kernels and offline use. We introduce
solver-based \emph{super-analysis}, a middle ground that uses expensive SMT-based analysis while transformations remain conventional, avoiding
program synthesis entirely. The key argument for super-analysis is that conventional compilers have rich transformation rules, but proving them to be safe requires complex reasoning: proving a safety check dead often needs facts from three
independent LLVM analyses \emph{simultaneously} --- a conjunctive entailment
problem suited to an SMT solver, not to any single dataflow pass.

We instantiate
super-analysis in \tool{}, an LLVM pass that proves sanitizer traps and
native-language safety checks unreachable and removes them.
One mechanism serves four frontend languages and yields speedups in each, including a $3.78\times$ speedup for a GEMM implemented in Julia.
A further solver-based optimization, \emph{guard synthesis}, handles
checks that are not dead for every input: the solver synthesizes a cheap invariant safety guard under which a fast check-free loop is cloned,
speeding up a Swift Base64 encoder by $1.44\times$ and a Julia LZ77
compressor by $4.01\times$. Compared with Souper, the synthesizing
superoptimizer for LLVM, \tool{} removes more safety checks in significantly less time, making it viable as an online compiler pass.
\end{abstract}
\ccsdesc[500]{Software and its engineering~Compilers}
\ccsdesc[300]{Theory of computation~Automated reasoning}
\keywords{SMT solvers, compiler optimization, sanitizers, bounds-check
elimination, undefined behavior, LLVM}
\maketitle
\section{Introduction}
\label{sec:intro}
Optimizing compilers decide how fast software runs. Every program in a
compiled language, from operating systems to an artificial intelligence training kernel, passes
through compiler optimization at an intermediate representation (IR) level. A compiler infrastructure such as
LLVM~\cite{lattner2004llvm} serves many input languages at once, including C and C++,
Swift, Rust, and Julia, all compiled to its IR and aggressively optimized through it. Almost all software nowadays goes through such optimization pipelines, so the effectiveness of optimizers heavily influences the performance of the software we use. 

Aggressive compiler
optimization currently falls into two general classes. \emph{Conventional optimizers},
such as LLVM's \texttt{-O3} pipeline, apply a fixed sequence of
analyses and rewrites. They are fast: they
make millions of decisions in milliseconds, each justified by a
heuristic or a lightweight analysis that reasons locally and gives up
early. They are perfect for online use in the programmer's develop-and-test loop. \emph{Superoptimizers}~\cite{massalin1987superoptimizer,
schkufza2013stoke, sasnauskas2017souper} are the extreme alternatives. They aggressively search for faster replacement
code and use a solver to prove each candidate correct. Correctness consists of checking if the candidate and the original piece of code are equivalent, usually offloaded to an SMT solver. Superoptimizers find more optimal code as they conduct an exhaustive search, in contrast with the conventional rule-based transformations.

Each class pays for its strength and has critical weaknesses. Conventional optimizers are fast, but they miss
optimizations whose safety needs deeper reasoning. Their lightweight conventional heuristics give up and are unable to provide sufficient analysis for many cases~\cite{taneja2020knownbits}. Superoptimizers are more aggressive in optimization as they find
code heuristics cannot, but their search is expensive and coupled with the solver's expensive check, as each candidate code found should be verified correct. Because of this, superoptimizers take up to hours to run, so they are confined to small code windows and run offline,
outside the everyday development and testing loops.
\begin{figure}[t]
\centering
\includegraphics[width=\linewidth]{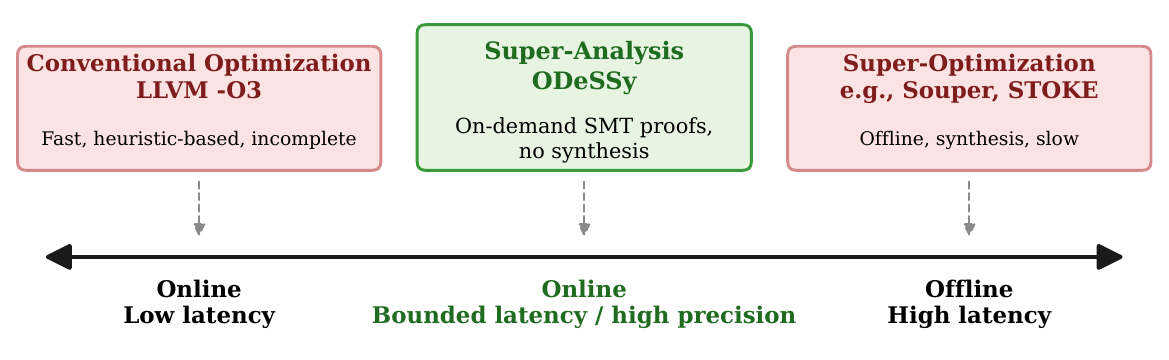}
\caption{The cost--precision spectrum of aggressive compiler
optimization. Conventional pipelines are online but heuristic and
incomplete; superoptimizers are precise but offline. Super-analysis
occupies the unpopulated middle: on-demand SMT proofs at bounded latency,
with no program synthesis, usable online.}
\Description{A horizontal axis from online/low-latency on the left to
offline/high-latency on the right. Three boxes sit along it: Conventional
Optimization (LLVM -O3) on the left, Super-Analysis (ODeSSy) in the
middle, and Superoptimization (Souper, STOKE) on the right.}
\label{fig:spectrum}
\end{figure}

This paper argues for an under-explored middle ground we call \emph{super-analysis}
(Figure~\ref{fig:spectrum}): use
the expensive solver-based analysis, and leave transformation
to the conventional machinery that already exists. No program is synthesized and
no search over the infinite space of programs is performed. The solver answers a minimal number of strategic questions, for example \emph{is this program point reachable under this context?}, and a
standard transformation (deleting an unreachable branch)
consumes the answer.

The term super-analysis was used by Lerner et al.~\cite{lerner2002composing} for different analyses combined into one lattice-based fixpoint, extending earlier hand-combined analyses~\cite{wegman1991constant, click1995combining}; we generalize it into a paradigm: expensive, unconventional analysis enabling conventional transformations. We introduce solver-based super-analysis as the most aggressive form, where an SMT solver is used as an analysis oracle (\S\ref{sec:related}).
We feed the solver with the formula $\mathit{Context} \wedge \mathit{Query}$, and ask for satisfiability. An
\textsc{unsat} verdict is a proof that licenses our transformations; a \textsc{sat} or \textsc{unknown}
verdict changes nothing.

SMT-based compiler analysis is more precise than native lightweight compiler analysis. Our central empirical finding is that the missed opportunities require complex reasoning over multiple domains and from different fact sources to prove safe. Proving
one bounds check dead in a real SHA-256 kernel
(Figure~\ref{fig:simple-example}) requires the return-value
range of one function, a scalar-evolution loop bound, a value-range fact
on a third value, and two dominating branch conditions from the program's prefix ---
\emph{simultaneously}. Each fact lives in a different LLVM analysis; no
single pass computes their conjunction, and the phase-ordering problem
poses an obstacle for a sequence of analyses to be successful. Deciding satisfiability of a
complex conjunction is, in principle, an entailment
problem --- exactly what an SMT solver decides. We inspect the unsat core of every proof, so this
claim is not an intuition but a measurement: the cores name the
contributing analyses, and they usually draw on several sources at once.

We instantiate super-analysis in \tool{}\footnote{On-Demand SMT System.},
an LLVM module pass targeting one of the most consequential classes of
provably-dead code in modern software: \emph{safety checks}. Sanitizers
such as UBSan~\cite{ubsanDocs} and the native checks of memory-safe
languages (Swift bounds and overflow checks, Rust panics, Julia bounds
errors) compile to trap-guarded branches
(Figure~\ref{fig:simple-example}) whose hot-path cost is the
standing tax of memory safety --- we measure it at up to $1.069\times$ on
the zstd library under signed-overflow checks and up to $4.15\times$ on a native-language kernel (Julia GEMM).
Today the only recourse to avoid this performance tax is to turn checks \emph{off}
wholesale (\texttt{-Ounchecked}, \texttt{--check-bounds=no}), trading away
the safety guarantee entirely. \tool{} offers the sound alternative:
delete exactly the checks that are provably redundant with the guards,
contracts, and invariants already present in the program, and keep every
check that is not.
\tool{} is able to recover up to $3.78\times$ of performance by removing redundant safety checks and enabling a cascade of further optimizations.

\begin{figure}[t]
\centering
\begin{minipage}[t]{\linewidth}
\begin{lstlisting}[style=odessy,
    morekeywords={var,for,in,let,if},
    comment={[l]{//}}]
var w = [UInt32](repeating: 0, count: 64)
for t in 16..<64 {
  // s0, s1: rotate-xors of w[t-15], w[t-2]
\end{lstlisting}
\begin{chkbox}{2}{40}
\begin{lstlisting}[style=odessy, xleftmargin=0pt,
    morekeywords={var,for,in,let,if},
    comment={[l]{//}}]
// implicit check (injected by compiler)
if t-16 >= w.count { trap() }
\end{lstlisting}
\end{chkbox}
\begin{lstlisting}[style=odessy,
    morekeywords={var,for,in,let,if},
    comment={[l]{//}}]
  w[t] = w[t-16] + s0 + w[t-7] + s1
}
\end{lstlisting}
\end{minipage}\par\vspace{0.7em}
\begin{minipage}[t]{\linewidth}
\begin{lstlisting}[style=odessy,
    morekeywords={load,add,icmp,uge,br,label,call,void,i8,i64,i1,ptr},
    comment={[l]{;}}]
; the check for w[t-16], after swiftc -O + LLVM O3:
%n   = load i64, ptr %w.count           ; array length: n = 64
%idx = add i64 %t, -16                  ; idx = t - 16
\end{lstlisting}
\begin{chkbox}{0}{58}
\begin{lstlisting}[style=odessy, xleftmargin=0pt,
    morekeywords={load,add,icmp,uge,br,label,call,void,i8,i64,i1,ptr},
    comment={[l]{;}}]
%oob = icmp uge i64 %idx, %n            ; oob = (idx >= n)
br i1 %oob, label %trap, label %ld      ; trap if oob
trap: call void @llvm.ubsantrap(i8 18)  ; aborts
\end{lstlisting}
\end{chkbox}
\end{minipage}
\caption{A Swift SHA-256 message-schedule loop (top) and one of its
bounds checks, which survives \texttt{swiftc -O} and LLVM O3, at the IR
level (bottom); checks highlighted. Conventional analysis cannot remove
the check from the hot loop, imposing a performance tax: scalar
evolution gives $t - 16 < 48$ and a return-value range gives
\texttt{w.count} $= 64$, but no analysis logically conjoins the two facts and
proves that the trap is unreachable.}
\Description{Above, a Swift listing of the SHA-256 message schedule
loop with the injected check, if t minus 16 is greater or equal to the
array count then trap, highlighted in a rounded pink box. Below, five lines of LLVM IR showing the
load of the array count, the index computation, and, highlighted in one rounded pink
box, the unsigned compare, the conditional branch to the trap block,
and the trap block's call to ubsantrap.}
\label{fig:simple-example}
\end{figure}

Deleting checks that are already dead is not \tool{}'s only
optimization. Many hot checks are not dead for every input: they are
dead only for the well-formed inputs a program actually receives, such
as arrays of the expected length. For these, \tool{} performs
\emph{guard synthesis}. The solver synthesizes a guard --- a cheap
condition over values that are invariant inside the loop --- under
which every check in the loop is provably dead. The loop is then
specialized: the guard is checked once, outside the loop, and when it
holds a copy of the loop runs with all its checks removed; otherwise
the original loop runs unchanged. Because real inputs take the
specialized copy, the hot loop becomes the fast one. Guard synthesis
stays within the paradigm: the solver synthesizes a condition, not a
program, and the transformation is an ordinary loop clone. On x86 it
speeds up a Swift Base64 encoder by $1.44\times$ and a Julia LZ77
compressor by $4.01\times$.

Super-analysis is a general compilation paradigm rather than a single
optimization, and it carries one open problem we deliberately defer:
\emph{when is expensive analysis worth invoking?} A solver consulted
indiscriminately at every program point would be ruinous, and deciding
where precision repays its cost demands a cost model no compiler has.
Safety-check elimination is the ideal first instantiation precisely
because it answers that question for free. The work list is given, and the traps are expected to be unreachable (therefore \textsc{unsat}) for safe programs. Based on our experiments, removing a safety check is either a win or has a marginal effect in most cases --- it simplifies the program control-flow, allowing other optimizations to be performed by subsequent passes. This
application therefore isolates the paradigm's contribution --- precision
of analysis --- from the question it defers: when to use it.

This paper makes the following contributions:
\begin{itemize}
\item \textbf{Super-analysis} (\S\ref{sec:superanalysis}): an aggressive compilation paradigm
  positioned between conventional heuristic-based optimization and superoptimization in terms of compiler latency and precision. The general recipe for this paradigm is to use solvers as an analysis oracle, while applying cheap conventional rule-based transformation for optimization. We identify heap reasoning as
  a crucial analysis problem, formalize it, and show that
  super-analysis can more effectively reason over heap and memory safety
  of certain transformations.
\item \textbf{Making super-analysis work} (\S\ref{sec:making}): a set of
  techniques inside the paradigm that make it effective. We encode the
  compiler's lightweight native analyses as facts that constrain the
  solver's search, and we
  introduce guard synthesis and loop specialization, which synthesize a
  runtime guard under which a loop's checks are dead and clone the loop
  behind it.
\item \textbf{\tool{}} (\S\ref{sec:system}): an LLVM pass and online optimizer
  as an instantiation of super-analysis. \tool{} eliminates safety checks and unlocks further optimizations. It supports C/C++, Swift, Rust, and Julia as frontends and can enable hot-path transformations and speedups.
\item \textbf{Robust Performance Improvement} (\S\ref{sec:results}):
  our systematic evaluation of \tool{} on an x86 server shows removing hot-loop checks has a cascading effect, unlocking further transformations (unrolling, vectorization, etc.) that enable up to $4.01\times$ speedups for widely used kernels and libraries. \tool{} proofs are generally cheap, which makes it usable as an online pass in the development loop. Multi-threading and per-query budgets make \tool{}'s analysis up to $2.4\times$ faster. A further comparison with Souper (a synthesizing superoptimizer) shows that \tool{} removes significantly more checks in less time.
\end{itemize}
\section{Paradigms of Aggressive Compiler Optimization}
\label{sec:related}
\begin{figure}[t]
\centering
\includegraphics[width=\linewidth]{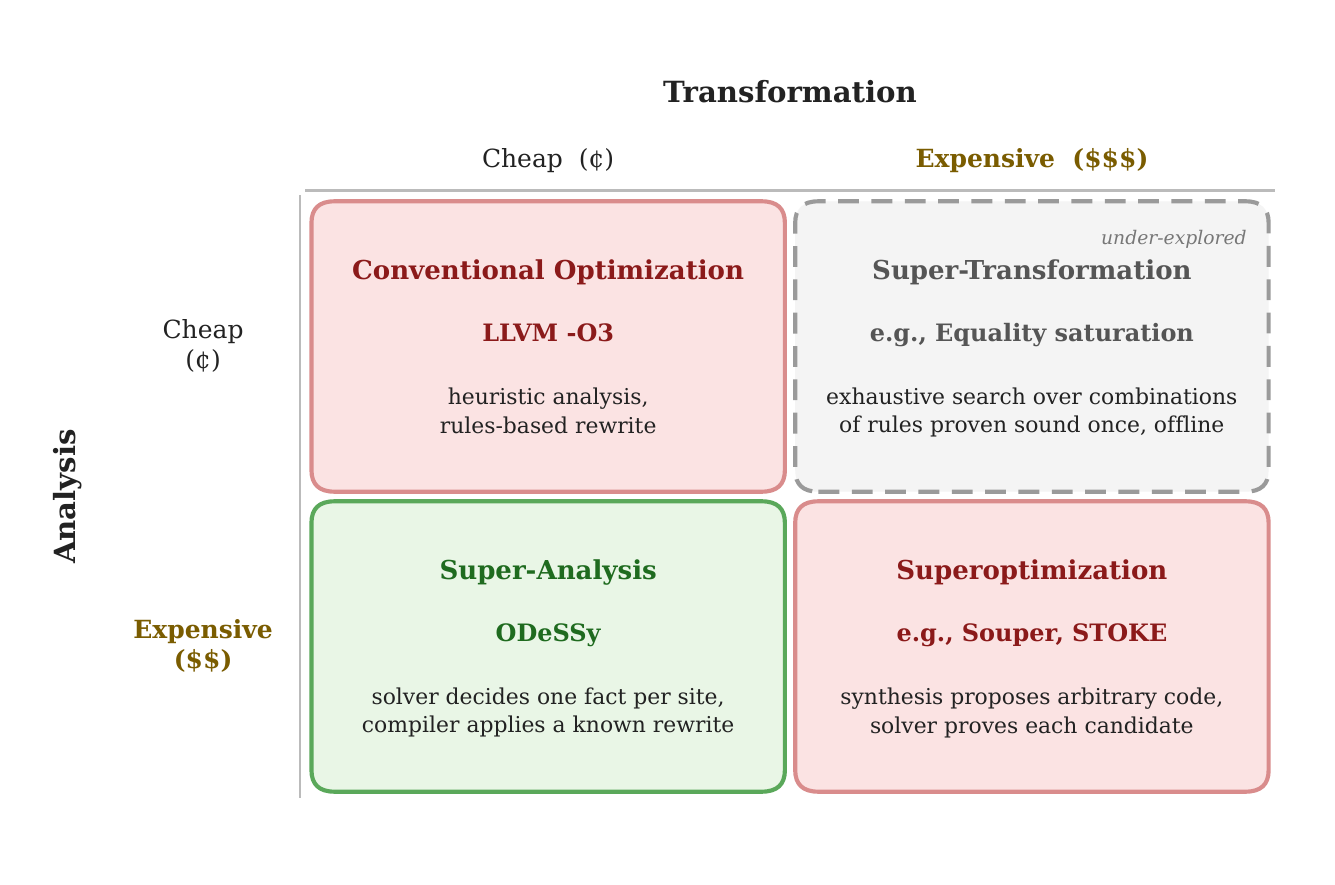}
\caption{Every optimizer is an
analysis that licenses a rewrite plus a transformation that performs it,
and each can be cheap or expensive. Conventional optimization keeps both
cheap; superoptimization makes both expensive. Super-analysis takes one
asymmetric corner where a solver licenses rewrites. Super-transformation takes the other: exhaustive search over
combinations of rewrites. Both asymmetric corners
are under-explored.}
\Description{A two-by-two matrix. Columns: transformation cost, cheap
versus expensive. Rows: analysis cost, cheap versus expensive. Cheap/cheap
is Conventional Optimization (LLVM -O3). Cheap analysis with expensive
transformation is Super-Transformation (e.g., equality saturation), marked
under-explored. Expensive analysis with cheap transformation is
Super-Analysis (ODeSSy). Expensive/expensive is Superoptimization (Souper,
STOKE).}
\label{fig:quadrants}
\end{figure}

Every optimizing pass consists of an analysis that licenses a rewrite and a
transformation that performs it, and each of the two can be cheap or
expensive. By taking the expensive or cheap route for each of the two components, we get to the different optimization families in Figure~\ref{fig:quadrants}, which include conventional optimization and superoptimization. We introduce two under-explored paradigms: (1) \emph{super-transformation}, of which equality saturation is a strict form, and (2) \emph{super-analysis}, a generalization of the combined analyses of Click and Cooper~\cite{click1995combining} and Lerner et al.~\cite{lerner2002composing}, and the focus of this paper. We systematically describe the design space these paradigms span, and how they differ and compare. 

\paragraph{Conventional optimization.} Conventional compiler optimization passes (like in LLVM) consist of analysis modules and transformation rules that consume the analysis to produce more efficient code. Both components are lightweight and fast, sacrificing precision to save compile time and provide rapid feedback. All the other families of optimizers deviate from this simple fast compilation methodology to perform more aggressive transformations.

\paragraph{Superoptimization.} A superoptimizer tries to search a large space of programs for a faster or shorter alternative to the initial code.
Massalin's original superoptimizer~\cite{massalin1987superoptimizer}
enumerated every instruction sequence up to a length bound and kept the
shortest one that matched the target on test inputs. Two lines of work
made the idea provably sound and practical. STOKE~\cite{schkufza2013stoke} replaces
enumeration with stochastic search: a random walk
mutates x86-64 kernels, scores each candidate by test-case correctness
and estimated latency, and hands the survivors to an SMT solver for a
formal equivalence proof, at costs of hours per loop-free kernel.
Souper~\cite{sasnauskas2017souper} works at the LLVM IR level: it
synthesizes replacement code for bounded peephole windows and verifies each
candidate with an SMT solver. In every sound superoptimizer the search proposes code and the solver
verifies it, so the solver is invoked \emph{per candidate}. 

Conventional optimization and superoptimization have been widely explored in the literature, while the asymmetric corners of Figure~\ref{fig:quadrants} are under-explored. We call the paradigm of cheap analysis and expensive transformation super-transformation, which is an exhaustive search over candidate transformations, coupled with cheap analysis. The corner that this paper extensively explores and instantiates is super-analysis: expensive (solver-based) analysis that enables cheap conventional transformations. We can state that superoptimization is a combination of super-transformation (synthesis as the most aggressive exhaustive search for alternative code) and super-analysis (solver-based equivalence checking in sound superoptimizers), making it super expensive.

\paragraph{Equality Saturation.}
Equality
saturation~\cite{tate2009equality} is a disciplined subset of super-transformation: rather than synthesizing candidates, it saturates an
e-graph with the combinations of a fixed rewrite-rule set and extracts
the cheapest result. This aggressive search ranges over combinations of rewrite rules rather than over
arbitrary programs, so no candidate needs an equivalence proof as the applied rules are sound and preserve equivalence.

\paragraph{Super-analysis.} This paper explores this fourth route. A
super-analyzer spends the solver on \emph{analysis} alone: the solver
decides whether a rewrite is licensed, and the compiler's existing conventional
transformations perform the rewrite. No replacement code is searched
for, synthesized, or verified. \S\ref{sec:superanalysis} develops the
paradigm. The combined analyses that Lerner et al.~\cite{lerner2002composing} call super-analyses belong to this paradigm in the broad sense: they combine several analyses into one, more precise than running them separately and more expensive than any one of them. When one analysis licenses a rewrite, the others reanalyze the rewritten code, which makes the combined analysis unconventional and more expensive, while the transformations it licenses remain conventional. They are at the cheap end of the paradigm. Logical products of abstract domains~\cite{gulwani2006combining} combine facts more directly. Our solver-based super-analysis is the most aggressive form, in which an SMT solver decides the conjunction of every analysis's facts, together with an exact encoding of the code. We use super-analysis to denote the aggressive, solver-based version from this point. 

\paragraph{How paradigms relate.} Our super-analysis methodology shares the verification engine with superoptimizers, though the difference in philosophy and architecture makes them radically different in terms of effectiveness and \emph{scaling}. Superoptimizers spend an expensive SMT equivalence-checking query per synthesized candidate, making them very slow. A super-analyzing compiler uses the expensive analysis more strategically, consulting the solver whenever its conventional analysis fails and the solver latency is worth the potential unlocked optimization. This makes super-analysis generally faster, something we show in \S\ref{sec:results-souper}. The unscalable nature of synthesis also limits it to small-window peephole optimizations, whereas a super-analyzer can reason more globally. For example, it can derive and consume post-conditions of complex loops and functions, which are beyond the scope of peephole optimizations.

\section{Super-Analysis}
\label{sec:superanalysis}

In this section we describe a general recipe for building a compiler super-analyzer.

\subsection{Encoding program context}
\label{sec:encoding}
A super-analyzer decides a property of a single \emph{program point} ---
here, whether a trap is reachable. The property is decided by
the \emph{context} that leads to the point.
The core machinery of the paradigm is therefore a translation: render
the context of a program point as a logical formula, and ask the solver
whether the property is consistent with it.

The translation proceeds \emph{backwards} from the point of interest.
In this paper we are concerned with safety checks and trap conditions, so starting at the trap condition, it walks the chain of definitions that
feed it and the branch conditions that dominate it, and renders each
instruction as an exact equation over bit-vectors --- arithmetic,
comparisons, shifts, overflow intrinsics, and the control flow
connecting them. The resulting formula is satisfiable exactly by the
machine states that can reach the point, so reachability becomes one
satisfiability query: $\mathit{Context} \wedge \mathit{TrapCond}$.

A backward encoding cannot descend forever. It stops at
\emph{boundaries} --- loop-carried values, function calls, loads:
boxes whose interior it does not open. What crosses a boundary is a
\emph{post-condition}: a fact true of the box's result in every
execution. The weakest post-condition is no constraint at all --- the
value is a free variable --- and this is the soundness floor: anything
not understood is over-approximated, so an \textsc{unsat} verdict holds
for every concrete behavior of the unencoded parts. From that floor,
post-condition derivation is a precision dial that can in principle be
turned arbitrarily far: a range, an algebraic relation among boundary
values, a loop invariant, a full summary. 

\subsection{Heap-Invariant Super-Analysis}
\label{sec:frame}

Encoding the context alone does not prove every property an optimizer
needs, because many of these properties concern the state of the heap:
optimizations depend on the values that loads return and on the
pointers they read through. A dynamic array's length, for example, is
stored as metadata in memory (in the array's header), so the length an
index is checked against is a load, and reasoning about it requires a
model of the heap. A compiler's scalar analyses do not provide one.
Each describes the values that one SSA register may hold: in LLVM,
value ranges (LazyValueInfo and range metadata), known bits, and the
trip counts and induction ranges of scalar evolution. They can
constrain the value of a single load, but they cannot relate two loads
to each other.

Figure~\ref{fig:frame-example} illustrates the need for heap-aware
reasoning. The example is a safety check in the inner loop of a matrix
multiply. Before the loop, a dimension guard loads \texttt{A}'s length
into $n_1$ and checks that it equals the loop bound $m$. Inside the
loop, the bounds check loads the length again, into $n_2$, and the loop
also stores into \texttt{C}. Since $i < m = n_1$, the check is dead if
$n_2 = n_1$, and scalar analyses cannot establish that, for two
reasons. First, a load is opaque to them: its result is whatever memory
holds, so without metadata $n_2$ may be any 64-bit value. Second,
$n_2 = n_1$ is a relation between two values, while a range or a set of
known bits describes one value at a time. The equality holds only if
the store between the two reads never overwrites \texttt{A}'s length,
which is a question about memory, answered in LLVM by alias analysis and
MemorySSA. Here they do not answer it: the reload is a copy made by the
vectorizer, without the aliasing metadata that would let alias analysis
rule out the store. The check survives \texttt{-O3}, and a solver given
only scalar facts treats $n_2$ as a free variable and finds the trap
reachable. The checks limit how much the loop is vectorized and
unrolled: in \texttt{generic\_matmatmul!}, the matrix multiply of
Julia's standard library, they cost $4.15\times$ on x86
(\S\ref{sec:frame-results}).

Heap-invariant super-analysis supplies the missing fact. A
super-analyzer can reason via axiomatic rules and encode the facts they
yield: here, that two loads through the same pointer return the same
value if no store in between overwrites the location --- an instance of
the frame rule of separation logic~\cite{reynolds2002separation}.
Super-analysis adds precision where conventional scalar analyses fail
(e.g., LLVM's known bits), and its special case, heap-invariant
super-analysis, takes over where conventional memory analysis fails
(e.g., LLVM's MemorySSA). A heap fact alone is rarely enough: the solver conjoins it with
the scalar facts --- for GEMM, the load equalities together with
trip-count, known-bits, and definitional facts --- and it is the
combination that proves the check dead. The encoding machinery that
handles GEMM (\S\ref{sec:frame-machinery}) also added proofs on other
benchmarks, such as the Swift and Julia SHA-256 implementations and
CryptoSwift.

\begin{figure}[t]
\centering
\begin{lstlisting}[style=odessy,
    morekeywords={load,store,icmp,ugt,ult,eq,br,label,i64,i1,ptr,x,double},
    comment={[l]{;}}]
; a hot loop after -O3 (schematic); A's length is in its header
%n1 = load i64, ptr %A.len             ; read 1, for the dimension guard
%ok = icmp eq i64 %n1, %m              ; size(A) == m, else throw
br i1 %ok, label %loop, label %throw
loop:                                  ; i runs from 0 to m-1
\end{lstlisting}
\begin{chkbox}{2}{82}
\begin{lstlisting}[style=odessy, xleftmargin=0pt,
    morekeywords={load,store,icmp,ugt,ult,eq,br,label,i64,i1,ptr,x,double},
    comment={[l]{;}}]
%n2 = load i64, ptr %A.len           ; read 2: the check reloads the array length
%oob = icmp uge i64 %i, %n2          ; the check needs i < n2 for each access to A
br i1 %oob, label %trap
\end{lstlisting}
\end{chkbox}
\begin{lstlisting}[style=odessy,
    morekeywords={load,store,icmp,ugt,ult,eq,br,label,i64,i1,ptr,x,double},
    comment={[l]{;}}]
  ;...
  store <2 x double> %v, ptr %C.data   ; can this store overwrite A.len?
  %c = icmp ult i64 %i.next, %m        ; back edge: bound is m
  br i1 %c, label %loop, label %exit
\end{lstlisting}
\caption{Two reads of one heap location with a store between them
(schematic LLVM IR of Julia's standard-library GEMM, check highlighted). The guard
gives $n_1 = m$ and the loop bounds $i$ by $m$, so the check is dead if
$n_2 = n_1$, which holds only if the store never overwrites \texttt{A}'s
length: a heap fact scalar analyses don't supply.}
\Description{A schematic LLVM IR listing: a load of A's length before
the loop, named n1, compared with the loop bound m by a dimension guard;
inside the loop, highlighted in one rounded pink box, a second load of
the same address, named n2, the compare of i against n2, and the branch
to the trap; a vector store into C's data, with a
comment asking whether it can overwrite A's length; and the back edge.}
\label{fig:frame-example}
\end{figure}
\subsection{Use Case: Safety Checks}
\label{sec:target}
We instantiate super-analysis on the \emph{safety checks} that compilers insert into
programs. A safety check guards an operation whose result would
otherwise be undefined or unsafe --- an array access out of bounds, a
signed integer overflow --- with a test and a branch to a block that
stops the program if the test fails. Every front end we support inserts
them. In C and C++, UndefinedBehaviorSanitizer~\cite{ubsanDocs} adds a
check for each kind of undefined behavior it is asked to catch; in trap
mode each one ends in a call to \texttt{llvm.ubsantrap}. Safer languages inject safety checks as part of their semantics. Swift checks
every array access and every arithmetic overflow by default, in
optimized builds too, and stops with \texttt{llvm.trap}~\cite{swiftBook}.
Rust checks every slice and array index, and every arithmetic operation
when overflow checks are enabled, and stops with a call to a panic
function~\cite{rustReference}. Julia checks every array access and stops
by throwing a bounds error~\cite{juliaBoundsCheck}. After lowering to
LLVM IR all of them have one shape --- a conditional branch to a block
that calls a function that never returns --- and that shape is what
\tool{} looks for (\S\ref{sec:pipeline}).

The checks make programs safe, and the safety has a price. In Julia's
standard-library matrix multiply, the check Julia inserts for the
access \texttt{A[i,l]} runs on every iteration of the innermost loop,
the hottest loop of the program, and as a
side exit from that loop it keeps LLVM from vectorizing and unrolling
the loop as fully as the check-free version: with the checks turned off
this kernel runs $4.15\times$ faster on x86. Today the only way to recover that performance
is to turn checks off wholesale (\texttt{@inbounds},
\texttt{--check-bounds=no}, \texttt{-Ounchecked}), giving up the guarantee
for the whole loop or program. \tool{} instead removes only the checks it
proves can never fail, so the program stays exactly as safe. Because
Swift and Rust keep their checks in optimized builds, \tool{} applies
to the large body of existing Swift and Rust code without any source
change --- two of our Swift benchmarks, crc32 and adler32, are taken
unmodified from existing code bases (\S\ref{sec:results-perf}) --- for
example as a step in release builds.

\section{Making Super-Analysis Work}
\label{sec:making}
\S\ref{sec:superanalysis} gave the general framework and recipe for super-analysis; this section collects the techniques that make
it practical and useful. Like
\S\ref{sec:superanalysis}, it is a general recipe, independent of
LLVM and applicable to any compiler. Encoding a program point's context may not provide enough constraints for a useful proof, therefore we import the lightweight facts a
conventional compiler analysis already computes, to further constrain the solver's search
(\S\ref{sec:conjunction}). Also, some checks
cannot be proven dead for every execution, no matter how precise the analysis is. We do not give up in these cases: we synthesize an invariant guard under
which checks are dead and clone the loop behind it (\S\ref{sec:guards}).

\subsection{Encoding Lightweight Heuristics}
\label{sec:conjunction}
Encoding the context alone does not produce every useful proof. The
encoding stops at boundaries, and a boundary value is free: the solver
may pick any value for it, including values no execution can produce.
Yet the compiler has usually learned something about that value
already. Its lightweight, fast native analyses --- value ranges, known
bits, loop trip counts --- run in milliseconds, and each describes a
slice of the truth. Importing their results as facts about the boundary
values constrains the solver's search space to executions that can
really happen. No single one of these facts is usually enough, though:
a proof needs several of them at once, and composing them is exactly
where a solver becomes necessary.

Consider again the simplified message-schedule loop of the Swift SHA-256
kernel of the introduction (Figure~\ref{fig:simple-example}), whose inner
16 iterations index a 64-element working array. Swift's bounds checks on
these accesses survive both \texttt{swiftc -O} and a subsequent full LLVM
O3. \tool{} proves them dead, and the extracted unsat cores show why no
single heuristic could. Each proof's minimal core has the shape:
\[
\textsf{Spec} \wedge \textsf{RM} \wedge \textsf{SCEV} \wedge \textsf{LVI}
\wedge \textsf{SCEV}' \wedge G_0 \wedge G_k \;\vdash\; \bot
\]
--- a return-value range (\textsf{RM}), two scalar-evolution facts
(\textsf{SCEV}), a lattice value range (\textsf{LVI}), and two dominating
guard conditions from the context ($G_0$, $G_k$), jointly inconsistent with the check
firing (\textsf{Spec}). Three independent analyses and path conditions must be intersected;
each in isolation is insufficient (removing any core member returns
\textsc{sat}). LLVM's analyses do not compose this way: each consumes the
IR and produces facts in its own domain, and no pass computes the
satisfiability of their conjunction --- nor could a fixed pipeline, since
which facts matter is different at every check site. Deciding such
conjunctions is the definitional competence of an SMT
solver~\cite{demoura2008z3}, making super-analysis necessary rather than merely convenient. Previous work has combined analyses in a single fixpoint over a product lattice~\cite{click1995combining, lerner2002composing}, whereas we use an SMT solver to directly conjoin the facts.

\subsection{Guard Synthesis and Loop Cloning}
\label{sec:guards}
Sometimes a check cannot be proved dead for every execution, because it
is not dead for every execution. Consider the Julia LZ77 compressor in our benchmarks. Its inner loop reads
\texttt{data[i+len]} only while \texttt{i+len <= n}, where $n$ is the
length of \texttt{data}, a value known only at run time. In
unbounded integer arithmetic this test makes the bounds check on
\texttt{data[i+len]} redundant. In 64-bit arithmetic it does not: if \texttt{n} is close to $2^{63}-1$, \texttt{i+len} can exceed $2^{63}-1$, the largest 64-bit integer. In this case Julia
integer arithmetic silently overflows, and the solver answers with a satisfying assignment in which \texttt{i+len} overflows to $-2^{63}$, passing the programmer's safety guard \texttt{i+len <= n} but triggering an out-of-bounds index (negative index) at \texttt{data[i+len]}. This vulnerability truly exists, and the safety check must remain for this array access.

Yet no real call looks like these models: an array of nearly $2^{63}$
bytes, about 9 exabytes, exists on no machine, and for every array that
does, every bounds check in the loop is dead. The property is hard to prove for the loop, but easy to prove
for a \emph{constrained} version of it --- and the constrained version is
the frequent one, often the only one that ever runs.

Guard synthesis exploits this. Instead of asking whether the check is
dead, we ask \emph{under what condition} it is dead: an invariant condition over
values that do not change inside the loop, cheap enough to check once
before the loop starts. Given such a guard, the loop is cloned
(Figure~\ref{fig:guard}). The fast copy has its checks removed; the
original, with every check intact, runs whenever the guard fails. For
\texttt{lz77\_scan}, guided by those models, the solver synthesizes
\[
0 \le n \le 2^{62},
\]
which holds for every array a real machine can allocate, so the fast
copy is the one that always executes.

\begin{figure}[t]
\centering
\begin{minipage}[c]{0.40\linewidth}
\begin{lstlisting}[style=odessy,comment={[l]{;}}]
Slow_Loop  ; with safety checks
\end{lstlisting}
\end{minipage}\hfill$\Longrightarrow$\hfill
\begin{minipage}[c]{0.50\linewidth}
\begin{lstlisting}[style=odessy,morekeywords={if,else},comment={[l]{;}}]
if Synthesized_guard:
    Fast_Loop  ; safety checks removed
else:
    Slow_Loop  ; checks in place
\end{lstlisting}
\end{minipage}
\caption{Guard synthesis and loop cloning. A loop whose safety checks
cannot be proved dead for every execution (left) is cloned behind a
synthesized guard (right). The guard is invariant w.r.t.\ the loop and is checked once.
When it holds, the fast copy runs with checks removed, and when
it fails, the original slow loop runs.}
\Description{On the left, a slow loop, commented as having safety
checks. An arrow points right, to an if-else: if the synthesized guard
holds, the fast loop runs, commented as having its safety checks
removed; otherwise the slow loop runs, commented as keeping its checks
in place.}
\label{fig:guard}
\end{figure}

The transformation itself is simple; the hard part is finding the right guard. It must
be a smart, useful guard: weak enough to hold on the executions we care
about, and loop-invariant and cheap, so that checking it at run time costs nothing
measurable. And it must be sufficient: under the guard, the fast copy
must be provably safe to run. Both are questions about the conjunction
of the guard, the context, and the trap condition, so, again, they
necessitate a solver. The solver proposes the guard, verifies that the
trap is unreachable under it, and verifies that the guard does not
contradict the context as a contradictory guard would make the fast
copy dead code.

The idea resembles abductive inference~\cite{dillig2012abductive,
dillig2013abductive}. Given known facts $\Gamma$ and a goal $\varphi$,
abduction seeks a hypothesis $\psi$ that is \emph{sufficient} and
\emph{consistent}:
\[
\Gamma \wedge \psi \models \varphi
\qquad\text{and}\qquad
\Gamma \wedge \psi \not\models \bot .
\]
Guard synthesis instantiates this directly. $\Gamma$ is the encoded
context of the check, with its dominating guards and imported facts; the
goal is that the trap is unreachable, $\varphi = \neg\,\mathit{TrapCond}$; and
the hypothesis is the guard $H$, restricted to values that do not change
inside the loop. Since $\Gamma \wedge H \models \neg\,\mathit{TrapCond}$
holds exactly when $\Gamma \wedge H \wedge \mathit{TrapCond}$ has no model,
each condition becomes one solver query:
\[
\underbrace{\Gamma \wedge H \wedge \mathit{TrapCond}\ \text{is \textsc{unsat}}}_{\text{sufficient: the fast copy is safe}}
\qquad\text{and}\qquad
\underbrace{\Gamma \wedge H\ \text{is \textsc{sat}}}_{\text{consistent: the guard is not vacuous}} .
\]
A guard that failed the second query would make the trap unreachable
only because the fast copy itself could never run. Among the guards that
pass both, we prefer the weakest, so that it holds on as many real
inputs as possible.

\section{\tool{}: Design and Implementation}
\label{sec:system}
\tool{} is an LLVM module pass of roughly five thousand lines that removes safety checks and enables further optimizations.
This section describes the pipeline the pass lives in
(\S\ref{sec:pipeline}), the encoding machinery at its core
(\S\ref{sec:encoding-impl}), inductive loop encoding
(\S\ref{sec:app-machinery}), the heap-invariance extension
(\S\ref{sec:frame-machinery}), guard synthesis and loop cloning
(\S\ref{sec:guards-impl}), and how proofs are audited before they are
believed (\S\ref{sec:audit}).

\subsection{Pipeline}
\label{sec:pipeline}
\tool{} runs sandwiched between two full \texttt{-O3} pipelines. The
first produces the optimized IR with inserted checks it could not prove dead; that is the analysis
input. The second, after \tool{}'s check elimination, re-optimizes the simplified
control flow and performs the transformations the deleted checks were
blocking --- unrolling, vectorization (\S\ref{sec:results-mech}). 

Internally the pass is three stages. \emph{Serial discovery} scans the
module for trap sites: calls to
\texttt{llvm.ubsantrap}/\texttt{llvm.trap}. Each incoming edge of a
trap block becomes one independent \emph{job}
(\S\ref{sec:encoding-impl}).
\emph{Parallel solve}: workers claim jobs in discovery order, each
owning a private Z3 context; the IR is read-only throughout. The only
shared mutable state is LLVM's own analyses, whose internal caches make
concurrent queries both a data race and a determinism hazard --- so
every analysis query passes through a ticket turnstile keyed on
discovery index, serializing them in one canonical order. That single
mechanism buys thread safety \emph{and} reproducibility: verdicts,
output IR, and logs are identical for \texttt{threads=1} and
\texttt{threads=N}, unless a query finishes close to its time budget, where
CPU contention can decide whether it completes (at 10\,s per query, five
Julia SHA-256 proofs and one of CryptoSwift's 379; \S\ref{sec:audit}). \emph{Serial
transform}: the main thread folds each proven branch to a constant in
discovery order --- the only IR mutation in the pass --- and standard
\texttt{simplifycfg}/\texttt{adce} reclaim the dead blocks. Workers
fence all exceptions and degrade to keeping the trap: every error path
in the system converges on declining to optimize.

\subsection{Encoding the context}
\label{sec:encoding-impl}
\label{sec:encoding-context}

\begin{table}[t]
\caption{The encoding rules. $\den{x}$ is the bit-vector term for SSA
value \texttt{x}, at \texttt{x}'s own width $w$; all arithmetic is
modulo $2^w$. $P(x)$ is a Boolean that holds when \texttt{x} is poison,
following LLVM's semantics of poison and \texttt{freeze}; each rule also sets
$P(z)$ to the disjunction of its operands' $P$.}
\label{tab:encoding}
\centering
\small
\resizebox{\linewidth}{!}{%
\begin{tabular}{l@{\hspace{1em}}l@{\hspace{1em}}p{0.3\linewidth}}
\toprule
LLVM IR (pseudocode) & encoding & also covers\\
\midrule
\texttt{z = add x, y} & $\den{z} = \den{x} + \den{y}$ & \texttt{sub}, \texttt{mul}\\
\texttt{z = udiv x, y} & $\den{z} = \den{x} \div_u \den{y}$ & \texttt{sdiv}, \texttt{urem}, \texttt{srem}\\
\texttt{z = shl x, y} & $\den{z} = \den{x} \ll \den{y}$ & \texttt{lshr}, \texttt{ashr}\\
\texttt{z = and x, y} & $\den{z} = \den{x} \mathbin{\&} \den{y}$ (Boolean when \texttt{i1}) & \texttt{or}, \texttt{xor}\\
\texttt{b = icmp ult x, y} & $\den{b} = \den{x} <_u \den{y}$ & all ten predicates\\
\texttt{z = zext x} & $\den{z} = \mathrm{zeroExtend}_w(\den{x})$ & \texttt{sext}, \texttt{trunc}\\
\texttt{z = select c, x, y} & $\den{z} = \mathrm{ite}(\den{c}, \den{x}, \den{y})$ & \\
\texttt{z = umax(x, y)} & $\den{z} = \mathrm{ite}(\den{x} >_u \den{y}, \den{x}, \den{y})$ & \texttt{umin}, \texttt{smax}, \texttt{smin}, \texttt{abs}, \texttt{fshl}\\
\texttt{r, o = sadd.with.overflow(x, y)} & $\den{r} = \den{x} + \den{y}$, \quad $\den{o} = \neg\,\mathit{NoOvf}_s(\den{x}, \den{y})$ & \texttt{\{s,u\} $\times$ \{add,sub,mul\}} \\
\texttt{z = add nsw x, y} & \texttt{add} rule, and $P(z) = P(x) \vee P(y) \vee \neg\mathit{NoOvf}_s(\den{x}, \den{y})$ & \texttt{nuw}; other poison flags\\
\texttt{z = freeze x} & $\den{z} = \mathrm{ite}\bigl(P(x),\ f,\ \den{x}\bigr)$, $f$ fresh;\quad $P(z) = \bot$ & \\
\texttt{z = phi [v$_1$, B$_1$], \ldots, [v$_n$, B$_n$]} & $\den{z} = \mathrm{ite}\bigl(\mathit{Reach}(B_n) \wedge \mathit{Edge}(B_n{\to}B),\ \den{v_n},\ \ldots\ \den{v_1}\bigr)$ & merge-point phis only\\
\texttt{br c, S, T} & $\mathit{Edge}(B{\to}S) = \den{c}$, \quad $\mathit{Edge}(B{\to}T) = \neg \den{c}$ & \texttt{switch}: case equalities\\
loop-header \texttt{phi}, \texttt{load}, \texttt{call}, \texttt{gep} & fresh variable & every unmodeled instruction\\
\bottomrule
\end{tabular}}
\end{table}

An encoding thread starts from a trap \emph{edge}, which is a boolean condition $c$ that takes the control-flow from a predecessor basic block $P$ to a trap block. Frontends merge error blocks, so a trap block routinely has many incoming
edges. Each qualifying edge is an independent obligation: the trap fires via this edge
iff control reaches $P$ and $c$ holds. Removing the exit branch from $P$ via $c$ to a shared trap block is sound
and profitable, even if the trap remains reachable from other blocks.

\begin{algorithm}[t]
\caption{Encoding and deciding one trap edge $(P, c)$}
\label{alg:encode}
\small
\begin{algorithmic}[1]
\State $G \gets \emptyset$ \Comment{dominating context}
\ForAll{dominators $D$ of $P$ ending in a cond.\ branch}
  \If{one outgoing edge of $D$ dominates $P$}
    $G \gets G \cup \{$that edge's condition$\}$
  \EndIf
\EndFor
\State $S \gets$ every instruction that $c$ and $G$ depend on, collected
  backwards through operands \Comment{the slice}
\Statex \hspace{\algorithmicindent} (including phis that merge branches; stopping at loads, calls,
  and loop phis)
\ForAll{$I \in S$ in reverse post-order}
  \State assert $\den{I}$ per Table~\ref{tab:encoding}
  \Comment{boundaries: fresh variables}
\EndFor
\State assert every $g \in G$ (tracked)
\State \Return $\mathrm{check\mbox{-}sat}(c)$
\Comment{\textsc{unsat}: fold edge; \textsc{sat}/\textsc{unknown}: keep}
\end{algorithmic}
\end{algorithm}

Encoding a job proceeds backwards from $c$ in three steps
(Algorithm~\ref{alg:encode}).
\emph{(1)~Guards.} Walk the dominator tree upward from $P$: for every
dominator ending in a conditional branch with one outgoing edge that
dominates $P$, every execution reaching $P$ took that edge, so the
edge's condition is asserted as context --- SSA immutability
guarantees its operands still denote the same values at the trap.
\emph{(2)~Slice.} A worklist walks operands backwards from $c$ and
from every guard. It stops at the \emph{boundaries} of
\S\ref{sec:encoding}: loads, address computations, calls we do not
encode, and loop-header phis, all of which become free variables. A
phi at an ordinary merge point is entered instead: its incoming values
and the branch conditions of its merge region join the slice. Every
cycle of data dependences in SSA runs through a loop-header phi, where
the walk stops, so the walk follows an acyclic graph and visits each value
once: it always terminates, and the slice is at most the size of the
function.
\emph{(3)~Translation.} The slice is encoded in reverse post-order by
the rules of Table~\ref{tab:encoding}: one exact bit-vector equation
per instruction, at the instruction's own width. A merge-point phi becomes
an if-then-else chain gated by a reachability formula that is built once per block:
\[
\mathit{Reach}(B) \;=\; \textstyle\bigvee_{P \in \mathrm{preds}(B)}
\mathit{Reach}(P) \wedge \mathit{Edge}(P{\to}B),
\]
The formula is an $O(\mathit{edges})$ circuit whose shared prefixes are stored once
(Z3 terms are hash-consed), leaving path enumeration to the SAT
engine rather than to an exponential DNF.

We encode LLVM's semantics of poison and
\texttt{freeze}~\cite{lee2017taming}, following
Alive2~\cite{lopes2021alive2}, which capture how undefined behavior
spreads through a program. Loaded values are assumed not to be poison: memory holds only values the
program stored, and poison can be stored only after undefined behavior that
the source languages either forbid in safe code (Rust, Swift) or check (our
UBSan specifications). The \texttt{poison-loads} knob drops this assumption
(Appendix~\ref{sec:app-defined-memory}).

The query is then $\mathit{Guards} \wedge \mathit{Definitions} \wedge
c$: satisfiable exactly by the machine states, consistent with
everything encoded, in which this edge reaches the trap.
\textsc{unsat} proves the edge dead in every execution;
\textsc{sat} and \textsc{unknown} (timeout) keep the check.

\subsection{Encoding LLVM's analyses}
\label{sec:system-facts}
The encoding of \S\ref{sec:encoding-impl} is the \emph{light} tier: it encodes the context alone and gives every boundary --- a load, a call, a loop-header phi --- an unconstrained variable. This over-approximates the values the boundaries can take, and the solver often answers \textsc{sat} with a model no execution can produce. The \emph{heavy} tier strengthens the
boundaries with post-conditions bought from LLVM's own analyses,
asserted as labeled constraints on the free variables:
\texttt{!range} metadata and range attributes (RM), KnownBits (KB),
LazyValueInfo ranges (LVI), scalar-evolution constant ranges (SCEV),
and a symbolic trip-count translation (SCEVSYM) that renders loop
bounds --- including symbolic starts and an exact unsigned-division
translation --- as constraints on loop-header phis. The values these
trip-count facts mention, such as the vectorizer's resume indices, are
themselves encoded by the rules of Table~\ref{tab:encoding} rather than
left free, and the bounds are stated in a form that is safe under
wrap-around (Appendix~\ref{sec:app-defs}). In debug mode
(\S\ref{sec:audit}) every fact is tracked under its label, so the unsat
core of each proof names the fact sources and mechanisms that contributed
to it; these are the labels in the cores of \S\ref{sec:conjunction} and
\S\ref{sec:frame-results}. 

Fast and Thorough (\S\ref{sec:results-cost}) combine these knobs. Besides these, the user sets the two latency dials, \texttt{timeout=} and \texttt{threads=} (\S\ref{sec:results-cost}); Appendix~\ref{sec:app-knobs} lists every knob.

\subsection{Encoding loops inductively}
\label{sec:app-machinery}
Hot loops hold the checks that cost the most, so we do not abandon such prime targets. We do not just rely
on LLVM's analyses either: we also encode constraints derived from the loop
itself. A loop-header phi merges a loop variable's value on entry with the
value its back edge carries from the previous iteration. Loop-header
phis are boundaries (\S\ref{sec:encoding}), so a check
whose safety depends on how a loop's values evolve is out of reach of a
single query unless a fact source bounds those values. Scalar evolution
often does (SCEVSYM), but not for loops that advance by a variable or
non-unit stride, or exit in several places. It is also not the most precise loop post-condition or invariant deriver.  \tool{}'s \texttt{ind} knob
proves such checks by one-step induction over the loop, with no trip
count and no invariant synthesis. Each loop-header phi has the form
\[
\phi = \mathtt{phi}\ [\phi_{\mathit{init}},\ \mathit{preheader}],\ [\phi_{\mathit{ind}},\ \mathit{BackEdge}],
\]
where $\phi_{\mathit{init}}$ is the value on entry to the loop and
$\phi_{\mathit{ind}}$ the value the back edge carries, computed in the
loop body. Let $\vec\phi_k$ be the vector of values of the trap's loop-header phis
in iteration $k$, and $\Gamma_k$ and $\mathit{TrapCond}_k$ the encoded
context, with its imported facts, and the trap condition in
iteration $k$; iterations $k$ and $k{+}1$ are encoded as two copies of
the loop body over separate variables, which in effect unrolls the loop
once. Two queries are posed:
\begin{align*}
\textsc{Base}:&\quad \Gamma_0 \wedge \vec\phi_0 = \vec\phi_{\mathit{init}} \wedge \mathit{TrapCond}_0\\
\textsc{Step}:&\quad \bigl(\Gamma_k \wedge \neg\,\mathit{TrapCond}_k \wedge \mathit{BackEdge}_k\bigr)
  && \text{iteration $k$: the check did not fire}\\
&\quad\wedge\ \vec\phi_{k+1} = \vec\phi_{\mathit{ind, k}}
  && \text{the back edge}\\
&\quad\wedge\ \bigl(\Gamma_{k+1} \wedge \mathit{TrapCond}_{k+1}\bigr)
  && \text{iteration $k{+}1$: the check fires}
\end{align*}
where $\mathit{BackEdge}_k$ states that iteration $k$ took the back
edge, $\vec\phi_{\mathit{ind, k}}$ is the back-edge value encoded in iteration $k$, and
$\neg\,\mathit{TrapCond}_k$ states that the check did not fire in
iteration $k$. If both are
\textsc{unsat}, the check cannot fire on the first iteration, nor on any
iteration that follows one on which it did not fire; by induction it
never fires. Iteration $k$ is
arbitrary, so the step query leaves $\vec\phi_k$ unconstrained except by
the facts below. \tool{} poses the two queries only when the direct query is
\textsc{sat} or \textsc{unknown} and the trap lies in a loop, first for
the trap's own loop and then for each enclosing one. Importing the facts of iteration $k$ queries scalar evolution, which is
not thread-safe, so the facts are imported through the turnstile of
\S\ref{sec:pipeline} and the two queries run in parallel like any other.

\subsection{Encoding heap invariance}
\label{sec:frame-machinery}
We add heap and pointer reasoning machinery to our pass. The \texttt{frame} encoding knob supplies the heap fact of \S\ref{sec:frame}:
two loads of the same location return the same value if no store
between them can write that location. Two notions from LLVM decide this.
A store (or call) \emph{clobbers} a load if it may write to the location
the load reads. \emph{MemorySSA} is LLVM's analysis that links each load
to the most recent memory write that may clobber it, using alias
analysis to skip writes that provably touch other locations.

For any two loads of the same pointer $p$, \texttt{n1 = load p} and
\texttt{n2 = load p}, where every path to \texttt{n2} passes through
\texttt{n1}, \tool{} adds the equality
\[
\frac{\mathrm{clobber}(n_2,\ p) \ \preceq\ \mathrm{def}(n_1)}
     {\den{n_1} = \den{n_2}}
\]
Read as an inference rule: $\mathrm{clobber}(n_2, p)$ is the most recent
write before $n_2$ that may clobber $p$, as MemorySSA reports it;
$\mathrm{def}(n_1)$ is the most recent write before $n_1$; and
$a \preceq b$ means that write instruction $a$ is $b$ itself or comes before it. The
premise therefore says that the last write that may change $p$ before
$n_2$ had already happened when $n_1$ read $p$. If the premise
holds, nothing between the two loads can change the location, and the
two loaded values are equal: a new encoded constraint. With no write
between the loads the premise holds trivially; with writes between them,
MemorySSA must show that none of them can write the location. This is exactly
the case of Figure~\ref{fig:frame-example}, where \texttt{n1} is read by
the dimension guard and \texttt{n2} by the check, and the equality
$n_2 = n_1$ is what proves the check dead.

\tool{} adds no alias analysis of
its own: every disjointness verdict is LLVM's, and when MemorySSA cannot
rule a write out, no fact is added. \tool{} only combines LLVM's analysis results with the other constraints, and that suffices for GEMM (Figure~\ref{fig:frame-example}): it recovers $3.78\times$ of the $4.15\times$ the checks cost (\S\ref{sec:frame-results}). A super-analyzer could reason itself about pointers, encoding address arithmetic and aliasing in the
solver, but more precise alias information does not by itself make code
faster: removing LLVM's alias analysis changes the final binaries little,
and only a small fraction of alias queries affects them~\cite{relaxingalias2025}.

\subsection{Implementing Guard Synthesis}
\label{sec:guards-impl}
\tool{} attempts guard synthesis (pass parameters \texttt{mv} and
\texttt{narrow}) only for a trap whose direct query returned
\textsc{sat}, and only when the trap sits inside a
loop. A trap that is already proved dead never reaches this step, so
guard synthesis can add proofs but never lose one. It proceeds in five
steps (Algorithm~\ref{alg:guards}): generate candidate guards, verify
that the trap is dead under them, shrink them to a small guard, check
the guard for vacuity, and transform the loop.

\begin{algorithm}[t]
\caption{Guard synthesis for one trap edge $(P, c)$ inside loop $L$}
\label{alg:guards}
\small
\begin{algorithmic}[1]
\Require the direct query $\Gamma \wedge c$ returned \textsc{sat}
  \Comment{$\Gamma$: context, facts, guards}
\State $C \gets$ candidates from templates T1--T3 over values
  invariant in $L$ \Comment{generation}
\If{$\mathrm{check\mbox{-}sat}(\Gamma \wedge c \wedge \bigwedge C) \neq
  \textsc{unsat}$} \Comment{verification}
  \State \Return \textit{keep the trap}
\EndIf
\State $H \gets C \cap \mathrm{unsat\mbox{-}core}$ of that query \Comment{a sufficient subset}
\ForAll{$h \in H$, upper bounds first, tightest first}
  \Comment{minimization}
  \If{$\mathrm{check\mbox{-}sat}(\Gamma \wedge c \wedge
    \bigwedge (H \setminus \{h\})) = \textsc{unsat}$}
    \State $H \gets H \setminus \{h\}$
  \EndIf
\EndFor
\ForAll{upper bounds $v \le_s 2^k$ in $H$} \Comment{loosening}
  \State raise $k$ while $\mathrm{check\mbox{-}sat}(\Gamma \wedge c \wedge \bigwedge H) = \textsc{unsat}$
\EndFor
\If{$\mathrm{check\mbox{-}sat}(\Gamma \wedge \bigwedge H) =
  \textsc{unsat}$} \Comment{vacuity check}
  \State \Return \textit{keep the trap}
\EndIf
\If{an operand of $H$ does not dominate the preheader of $L$}
  \State \Return \textit{keep the trap}
\EndIf
\State clone $L$ into $L_{\mathit{fast}}$; in the preheader, branch on
  $\bigwedge H$ \Comment{transformation}
\Statex \hspace{\algorithmicindent} to $L_{\mathit{fast}}$ if it holds,
  to the original $L$ otherwise
\State in $L_{\mathit{fast}}$, fold the trap edge to its non-trap
  successor
\end{algorithmic}
\end{algorithm}

\paragraph{Templates.} Every candidate is a comparison over values that
do not change inside the loop, so it can be evaluated once in the loop's
preheader. Three templates generate the candidates.
\begin{itemize}
\item \textbf{T1, length versus index.} When the trap condition is
  $\mathit{idx} \ge_u \mathit{count}$, with $\mathit{count}$
  loop-invariant and $\mathit{idx}$ varying, propose
  $\mathit{count} >_u \mathit{hi}$, where $\mathit{hi}$ is the largest
  value the index can take according to its static unsigned range.
\item \textbf{T2, sane range.} For every loop-invariant integer input
  $v$, propose $0 \le_s v$ and a ladder of upper bounds $v \le_s 2^k$:
  far above realistic sizes, far below overflow.
\item \textbf{T3, symbolic index bound.} When the index is an affine
  function of the loop counters, compute its largest value over the
  whole loop nest from the loops' trip counts, and propose
  $\mathit{count} >_u \max(\mathit{idx})$. For \texttt{matmul!} this
  yields $|a| > n^2 - 1$.
\end{itemize}

\paragraph{Verification.} All candidates are asserted together, each
tracked, on top of the context, the facts, and the trap condition. If
the solver answers \textsc{unsat}, the trap is dead under their
conjunction, and the unsat core names a sufficient subset $H$. If it
answers \textsc{sat} or \textsc{unknown}, no guard the templates can
express suffices, and the trap is kept.

\paragraph{Minimization and loosening.} Cores are not minimal. We drop
the conjuncts of $H$ one at a time, upper bounds first and tightest
first, keeping every drop that stays \textsc{unsat}. We then try to
relax each remaining upper bound, for example from $2^{15}$ to $2^{30}$,
while the query stays \textsc{unsat}. The result is a guard that holds
on as many real inputs as the proof allows.

\paragraph{Vacuity.} A guard that contradicts the context makes the trap
trivially unreachable, and the fast copy dead code. We therefore check
that $\Gamma \wedge H$ is satisfiable, and refuse the guard otherwise.

\paragraph{Transformation.} Every operand of $H$ must dominate the
loop's preheader, so that $H$ can be evaluated there. The loop is
cloned; a branch on $H$ in the preheader selects the fast copy or the
original; in the fast copy, the proven trap branches are folded. Nested
loops are handled innermost first, each guard is hoisted to the
outermost loop in which it is invariant, and later clones inherit the
folds already made in earlier ones. When a verification query contains
64-bit multiplications that the solver struggles with, it is re-encoded
at 32 bits with explicit overflow flags; this is sound only under $H$,
which bounds the inputs, so it is used only to certify guards.

\subsection{Caching for Incremental Compilation}
\label{sec:cache}
To make \tool{}'s feedback faster so that it can be used more conveniently in the programmer's develop-and-test loop, we cache the solver's verdict for each compiled function. This feature is useful as programmers usually implement modules incrementally, where each compile sees similar code with small changes. \tool{}
therefore keeps an optional per-function \emph{verdict cache}. This feature is off by default.

The cache is keyed by the SHA-256 hash over a normalized form of the function (value and block names
numbered, debug information dropped, metadata and attributes compared by
content), over what the function sees outside itself (its callees'
declarations and attributes, the globals it references, the data layout and
target), and over the configuration that can change a decision (the pass
parameters, \tool{}'s own version, and the LLVM and Z3 versions). The cache
stores \tool{}'s decisions for the function: which checks
are removed, and which loops are cloned under which guards. On a hit,
\tool{} skips the analyses, the encoding, and the solver for that function
and replays the stored decisions through the same transformation code.

\subsection{Debug Mode}
\label{sec:audit}
We take multiple measures to ensure the absence of data races,
parallelism bugs, and encoding bugs. These checks run in a debug mode as each one would impose a performance tax on the pass. First,
verdicts, output IR, and logs must be identical for \texttt{threads=1}
and \texttt{threads=N} up to the time-budget exception of \S\ref{sec:pipeline}, which we verify by \texttt{diff}.

To make sure our tool is sound, in debug mode no
\textsc{unsat} is accepted at face value; each one is inspected. We perform a \texttt{vacuity} test, meaning that we remove the trap condition from the \textsc{unsat} proof and query the solver again with all the other constraints. If
the context alone is already unsatisfiable --- an encoder bug, or a
genuinely contradictory guard chain --- the elimination is
\emph{refused} and reported. Vacuous counts are reported in every static
audit and are zero in the vast majority of cases, so a vacuous proof can be
considered a warning. The timed builds run without this audit to avoid an extra SMT query, for speed; guard
synthesis checks its own guards for vacuity in every build
(\S\ref{sec:guards-impl}).

Independently, every encoded solver constraint is tracked, so each
\textsc{unsat} yields a minimal core naming exactly the guards and
facts that proved it; we inspect the cores of every nontrivial proof
family by hand before counting it.

\section{Experimental Setup}
\label{sec:results}
\subsection{Experimental Methodology}
\label{sec:method}
\paragraph{Machines.}
x86 measurements use a CloudLab~\cite{cloudlab} \texttt{c220g2} node:
dual Xeon E5-2660~v3 (Haswell-EP, 10 cores/socket, AVX2, no AVX-512),
DDR4-2133, Ubuntu 24.04, with Intel turbo disabled and every timed run
pinned to socket~0 (\texttt{numactl}). Apple-Silicon measurements use an
M-series MacBook Pro. The pass builds against pinned LLVM trunk
(23.0.0git) with Z3 4.8.12; Swift 6.3.3, Rust 1.97, Julia 1.12.
\paragraph{Timing protocol.}
To measure the run time of our compiled benchmarks, we run each output binary 30 times and report the median to rule out outliers. We compare
against two baselines. The first runs \texttt{-O3} once. The second runs
it twice in a row, as \tool{} itself sits between two \texttt{-O3} runs
(\S\ref{sec:pipeline}), so a speedup over it cannot come from the extra
\texttt{-O3} run alone and is \tool{}'s own contribution. The two
baselines perform almost identically, and the difference between them
serves as the noise floor of each run. Speedups are reported against the
single \texttt{-O3} baseline; against the double one, every speedup
changes by at most 1.1\%. The runtime experiments (\S\ref{sec:results-perf}--\S\ref{sec:results-mech}) use a 10\,s per-query solver budget; the compile-cost experiments (\S\ref{sec:results-cost}) vary it.

\paragraph{Benchmarks, specifications, tiers.}
Two C libraries (zlib, zstd) under UBSan trap-mode
specifications; eight Swift kernels plus the unmodified CryptoSwift
library; Julia kernels including the stdlib GEMM of \S\ref{sec:frame};
and Rust ports that exercise the \texttt{traps=} retargeting. Every x86 runtime experiment uses one
configuration: the analysis facts and frame facts of the full tier, plus
inductive encoding and guard synthesis
(\texttt{heavy;frame;ind;mv;narrow}, 10\,s per query).
\paragraph{Research questions.}
Our evaluation is organized around four questions:
\begin{itemize}
\item \textbf{RQ1 --- How much does \tool{} improve performance?}
  How many checks can \tool{} prove dead, and --- where proofs land on
  hot paths --- how much speedup does it gain? (\S\ref{sec:results-perf})
\item \textbf{RQ2 --- What is the source of speedup?}
  Which transformations do hot-path proofs unlock, and which component
  of \tool{} makes guard synthesis pay? (\S\ref{sec:results-mech})
\item \textbf{RQ3 --- Is \tool{} fast enough for real-world use?}
  What does \tool{} cost to compile with, and how far can its budget be
  turned down? (\S\ref{sec:results-cost})
\item \textbf{RQ4 --- How does super-analysis compare with superoptimization?}
  On the same inputs, what does each paradigm remove, and at what compile
  time? (\S\ref{sec:results-souper})
\end{itemize}
\subsection{RQ1: How much does \tool{} improve performance?}
\label{sec:results-perf}
\label{sec:results-static}

In this section we evaluate how much \tool{}'s proofs speed up the final compiled binary. We collected 17 benchmarks, 6 of them from existing libraries and repositories or extracted from them, and the rest are our implementations of memory- and compute-heavy everyday kernels. Table~\ref{tab:static} lists the benchmarks, their sources, and the percentage of their checks that \tool{} removes. 

What matters more than the static percentage of check removals in Table~\ref{tab:static} is how much performance improvement each final compiled binary yields, in other words whether check removal buys performance, which Table~\ref{tab:speedups} establishes. We achieve a significant speedup of $4.01\times$ on Julia lz77
and $3.78\times$ on stdlib GEMM; in total 7 out of 17 benchmarks have a performance improvement of more than 10\%. These performance improvements span different applications: matrix operations are widely used in scientific computing and machine learning, and so are compression libraries and cryptographic kernels. A 10 to 300\% performance improvement is very impactful, especially knowing that code safety is not sacrificed. Also, for all four of our frontend languages (C, Swift, Rust, Julia) we have benchmarks that speed up, suggesting that \tool{} is not bound to a particular frontend or application, and super-analysis can be applied to a variety of domains. The results carry over to another architecture. On Apple M-series \tool{} recovers performance by similar magnitudes: $4.14\times$ on GEMM, $2.60\times$ on Julia lz77, and $1.146\times$ on zstd compression (Appendix~\ref{sec:app-mseries}).

The safety check performance tax, or as we call it, the ceiling, is how much the checks slow the code down. Ideally \tool{} would recover all of the performance loss, and it recovers nearly all of it for 5 benchmarks. For four benchmarks (Rust lz77, the Swift utf8 validator, Swift sha256, and md5) \tool{}'s partial check removal yields better performance than full removal (higher than ceiling). We have three interesting cases where check removal slows down the code, and one (CryptoSwift) where despite having a positive ceiling, \tool{}'s check removal slows the computation. Removing a check that blocks no further transformation leaves the second \texttt{-O3} free to make different code-generation choices, which can cost as well as gain (\S\ref{sec:results-mech}).

\begin{table}
\caption{The benchmarks and the checks \tool{} proves dead in each.
``Source'': \emph{Repo} = an unmodified open-source repository;
\emph{Ours} = our implementation of a standard kernel. Proven: the
percentage of the checks left after \texttt{-O3} that \tool{} removes,
including those removed in guarded loop copies (\S\ref{sec:guards}) (x86, Thorough, 10\,s per query; the same builds
as Table~\ref{tab:speedups}). A check is one trap edge
(\S\ref{sec:encoding-impl}); several edges can share one trap, so fewer
traps than edges disappear from the final binary.}
\label{tab:static}
\small
\begin{tabular}{lllr}
\toprule
Benchmark & Lang & Source & Proven\\
\midrule
zlib & C & Repo & 19\%\\
zstd & C & Repo & 25\%\\
CryptoSwift & Swift & Repo & 14\%\\
crc32 & Swift & Repo & 30\%\\
adler32 & Swift & Repo & 5\%\\
sha256 & Swift & Ours & 22\%\\
lz77 & Swift & Ours & 26\%\\
base64 & Swift & Ours & 26\%\\
utf8 validator & Swift & Ours & 19\%\\
sha1 & Swift & Ours & 8\%\\
md5 & Swift & Ours & 4\%\\
lz77 & Rust & Ours & 50\%\\
matmul & Rust & Ours & 50\%\\
gemm & Julia & Repo & 56\%\\
sha256 & Julia & Ours & 81\%\\
matmul & Julia & Ours & 100\%\\
lz77 & Julia & Ours & 50\%\\
\bottomrule
\end{tabular}
\end{table}

\begin{table}
\caption{Runtime results on x86. Speedup = median checked time / median
\tool{} time; ceiling = median checked time / median unchecked time; the
checked program goes through the same \texttt{-O3} pipeline. Medians of 30
shuffled repetitions (21 for the Julia rows), byte-identical output across
all builds.}
\label{tab:speedups}
\small
\begin{tabular}{llrr}
\toprule
Benchmark & Front end & Speedup & Ceiling\\
\midrule
lz77$^{a,b}$ & Julia & $\mathbf{4.013\times}$ & $4.013\times$\\
gemm, $n{=}512^{a,c}$ & Julia & $\mathbf{3.778\times}$ & $4.154\times$\\
matmul & Rust & $\mathbf{1.453\times}$ & $1.507\times$\\
lz77$^{a}$ & Julia & $\mathbf{1.45\times}$ & $4.013\times$\\
base64 & Swift & $\mathbf{1.436\times}$ & $1.430\times$\\
matmul, $n{=}512^{a}$ & Julia & $\mathbf{1.128\times}$ & $1.129\times$\\
sha256$^{a}$ & Julia & $\mathbf{1.107\times}$ & $1.128\times$\\
sha256 & Swift & $\mathbf{1.105\times}$ & $1.093\times$\\
crc32 & Swift & $\mathbf{1.074\times}$ & $1.074\times$\\
zstd, compression$^{c}$ & C & $\mathbf{1.057\times}$ & $1.069\times$\\
adler32 & Swift & $\mathbf{1.039\times}$ & $1.137\times$\\
lz77 & Rust & $\mathbf{1.030\times}$ & $0.998\times$\\
utf8 validator & Swift & $\mathbf{1.025\times}$ & $0.969\times$\\
zlib & C & $\mathbf{1.011\times}$ & $1.054\times$\\
\midrule
md5 & Swift & $\mathbf{0.983\times}$ & $0.815\times$\\
CryptoSwift & Swift & $\mathbf{0.972\times}$ & $1.113\times$\\
sha1 & Swift & $\mathbf{0.956\times}$ & $0.975\times$\\
lz77 & Swift & $\mathbf{0.830\times}$ & $0.830\times$\\
\bottomrule
\end{tabular}\\
\raggedright\footnotesize
$^{a}$Julia is JIT-compiled and cannot consume ahead-of-time LLVM IR, so
each Julia row times a separate copy of the unchanged kernel, transformed
by hand exactly as \tool{}'s proofs license (\texttt{@inbounds} on the
proven checks, or loop versioning under the synthesized guard).
$^{b}$Guard synthesized under a 60\,s solver budget; at the default
10\,s budget, one of the two loads' checks is proven (Julia lz77 below).
$^{c}$zstd: median of five independent builds (range
$1.018$--$1.059\times$); gemm: median of three Julia processes (range
$3.60$--$3.97\times$).\par
\end{table}
\subsection{RQ2: What is the source of speedup?}
\label{sec:results-mech}
In this section we explain how proofs unlock massive speedups, and in our main ablation study we explore which components of our guard synthesis procedure mainly contribute to performance improvement. 

A removed check pays exactly when it was blocking a further
transformation. A check is a side exit; inside a hot loop it prevents
unrolling and vectorization, and once it is gone the second
\texttt{-O3} performs them, so a few proofs can cascade into a large
speedup. SHA-256 is the typical case: seven proofs in the
message-schedule loop let it unroll, and the kernel runs $1.105\times$
faster on x86. Where the proofs land
matters more than how many there are: on zlib the proofs remove 208
traps from the final binary, all on cold paths, and leave it flat
($1.011\times$), while on zstd they remove only 13 and buy $1.057\times$
of compression's $1.069\times$ ceiling.

\paragraph{Heap invariance.}
\label{sec:frame-results}
Heap invariance unlocks stdlib GEMM, the second-largest speedup in
Table~\ref{tab:speedups}. Its innermost loop, the hottest code of the
kernel, makes four array accesses per iteration: it reads \texttt{A[i,l]},
\texttt{B[l,j]}, and \texttt{C[i,j]}, and writes \texttt{C[i,j]}, and Julia
guards each with a bounds check that reloads the array's length from its
header (Figure~\ref{fig:frame-example}). With the scalar facts alone,
\tool{} proves none of these checks: each reload is a fresh variable. With
the frame facts of \S\ref{sec:frame-machinery}, each reload is equated
with the load of the same length before the loop, where Julia's own
dimension guard compares it with the loop bound, and \tool{} proves 9 of
the kernel's 16 trap edges, among them every check of the read of
\texttt{A} and of the write of \texttt{C}. The proof of the read's check
draws on six fact sources at once
($|\mathrm{SCEVSYM}|\,|\mathrm{KB}|\,|\mathrm{SCEVEQ}|{\times}3\,
|\mathrm{FRAME}|{\times}3$, plus four guards).
These two accesses are the ones that matter. Each check is a side exit
from the innermost loop, and with the read of \texttt{A} and the write of
\texttt{C} unchecked, LLVM aggressively unrolls and vectorizes the loop as it does the
check-free version. Julia's JIT cannot consume the pass output, so we
annotate exactly the two proven accesses \texttt{@inbounds} and leave the
reads of \texttt{B} and \texttt{C} checked: the kernel runs $3.778\times$
faster on x86, of the $4.154\times$ that removing all four checks gives.
The speedup is robust and consistent across matrix dimensions: \tool{}'s
build is faster than the checked kernel at every one of 14 problem shapes,
by $1.5$--$4.1\times$ (Figure~\ref{fig:gemm-sweep}).

\paragraph{Guard synthesis: which component pays?}
Most of the other large speedups --- Julia lz77 ($4.01\times$), base64
($1.44\times$), Rust matmul ($1.45\times$), Julia matmul, crc32 --- come
from guard synthesis and loop cloning (\S\ref{sec:guards}). We are interested in knowing if our synthesis algorithm makes the difference, or \tool{}'s verification and proving ability is again buying us the performance. Therefore we isolate the contribution of each component in this experiment.

Cloning a
loop behind a runtime test is not new (\S\ref{sec:relwork}): LLVM's own
inductive range-check elimination (IRCE) splits a loop around its range
checks and versions it behind a runtime range test. Using IRCE we can answer the question: is it our smart guards or is it \tool{}'s proving ability that makes the difference? We
separate the two components with five pipelines that combine the two parts with
LLVM's counterparts:
(i)~\tool{} with guard synthesis: \tool{}'s guard and \tool{}'s proofs;
(ii)~\tool{}'s guard and loop copy, with every check left in place, so
that \texttt{-O3} must remove them: \tool{}'s guard, LLVM's proofs;
(iii)~IRCE followed by \tool{} without guard synthesis: LLVM's guard,
\tool{}'s proofs;
(iv)~IRCE followed by \texttt{-O3}: LLVM's guard and proofs; and
(v)~\texttt{-O3} alone. 
Table~\ref{tab:ablation} reports proofs and speedups on the six
benchmarks whose speedup involves guard synthesis.
\begin{table}
\caption{Guard-synthesis ablation. Each
pipeline has two columns: the speedup over the unchanged checked
program, and the number of checks it proves dead. Pipeline (i) is the best or tied on every benchmark. Pipelines (ii) and (iii), which use only one \tool{} component, are better than or equal to (iv) and (v) in most cases.}
\label{tab:ablation}
\small
\setlength{\tabcolsep}{3pt}
\resizebox{\linewidth}{!}{%
\begin{tabular}{ll*{5}{rr}}
\toprule
 & & \multicolumn{2}{c}{(i) \tool{} guard} & \multicolumn{2}{c}{(ii) \tool{} guard} & \multicolumn{2}{c}{(iii) IRCE guard} & \multicolumn{2}{c}{(iv) IRCE guard} & \multicolumn{2}{c}{(v) no guard}\\
 & & \multicolumn{2}{c}{+ \tool{} proofs} & \multicolumn{2}{c}{+ \texttt{-O3}} & \multicolumn{2}{c}{+ \tool{} proofs} & \multicolumn{2}{c}{+ \texttt{-O3}} & \multicolumn{2}{c}{\texttt{-O3}}\\
\cmidrule(lr){3-4}\cmidrule(lr){5-6}\cmidrule(lr){7-8}\cmidrule(lr){9-10}\cmidrule(lr){11-12}
Benchmark & Front end & Speedup & Proofs & Speedup & Proofs & Speedup & Proofs & Speedup & Proofs & Speedup & Proofs\\
\midrule
lz77$^{a}$ & Julia & $\mathbf{4.013\times}$ & \textbf{4} & $1.000\times$ & 0 & $1.000\times$ & 0 & $1.000\times$ & 0 & $1.000\times$ & 0\\
base64 & Swift & $\mathbf{1.435\times}$ & \textbf{8} & $1.259\times$ & 7 & $1.248\times$ & 3 & $1.243\times$ & 3 & $1.000\times$ & 1\\
matmul$^{c}$ & Rust & $\mathbf{1.281\times}$ & \textbf{2} & $1.279\times$ & 2 & $1.173\times$ & 2 & $1.174\times$ & 2 & $1.000\times$ & 0\\
matmul, $n{=}512$ & Julia & $\mathbf{1.128\times}$ & \textbf{3} & $1.000\times$ & 0 & $1.000\times$ & 0 & $1.000\times$ & 0 & $1.000\times$ & 0\\
sha256$^{b}$ & Julia & $\mathbf{1.089\times}$ & \textbf{13} & $1.021\times$ & 7 & $1.031\times$ & 7 & $1.024\times$ & 2 & $1.024\times$ & 2\\
crc32 & Swift & $\mathbf{1.074\times}$ & \textbf{11} & $1.071\times$ & 4 & $1.070\times$ & 4 & $1.000\times$ & 0 & $1.000\times$ & 0\\
\bottomrule
\end{tabular}}\\
\raggedright\footnotesize
$^{a}$Guard synthesized under a 60\,s budget; at 10\,s, (i) removes 2
of the 4 checks. $^{b}$At the 10\,s budget (i) leaves one access checked; the full
guarded copy of Table~\ref{tab:speedups} reaches $1.107\times$.
$^{c}$Rust matmul varies between runs: the same build measures between
$1.28\times$ and $1.47\times$ across our runs ($1.453\times$ in
Table~\ref{tab:speedups}). The five pipelines here are timed in one run, so they
are comparable with each other.\par
\end{table}

\emph{Pipeline (i) is the best or tied on every benchmark}, demonstrating that the combination of guard synthesis and \tool{} proofs is the most powerful and fruitful. Pipeline (i) either matches or beats every other pipeline. It is the only pipeline powerful enough to recover the $4.013\times$ of Julia lz77. 

Using guard synthesis or \tool{} proofs alone still recovers a lot of performance and yields a lot of proofs, although it is hard to say exactly which of the two, (ii) or (iii), is better. Both of them are better than or equal to the full LLVM pipelines (iv) and (v) in most cases. \emph{This restates the main thesis of the paper: solver-based expensive analysis pays back.} We believe our guard synthesis algorithm and templates are powerful, though it beats LLVM's IRCE not just because of our smart design choices, but \emph{because our guard synthesis is solver-guided from the beginning.}

\subsection{RQ3: Is \tool{} fast enough for real-world use?}
\label{sec:results-cost}
\paragraph{Compile cost.}
Compile-time latency is the first concern when it comes to SMT-assisted optimization, and it has limited previous work to offline use~\cite{sasnauskas2017souper, schkufza2013stoke, taneja2020knownbits}. A compiler pass must be fast to be widely adopted by programmers as an online pass, and we use three measures to make \tool{} faster: massive parallelism, strict SMT timeouts, and relaxed encoding precision. This makes \tool{}'s cost flexible, and our results show that it is suitable for online use inside the programmer's development loop.

To make \tool{} easy and simple to use, we introduce two tiers it can run in: \emph{Fast} and \emph{Thorough}. Fast uses strict 10\,ms timeouts with 10 threads, and encodes the imported LLVM analyses (\S\ref{sec:system-facts}), heap invariance (\S\ref{sec:frame-machinery}), and guard synthesis (\S\ref{sec:guards-impl}), without inductive encoding. Thorough is the configuration behind the speedup table (Table~\ref{tab:speedups}) and optimizes more aggressively, using a generous 10\,s timeout, 10-thread parallelism, and all the fruitful knobs on: the imported analyses, heap invariance, inductive encoding (\S\ref{sec:app-machinery}), and guard synthesis (Appendix~\ref{sec:app-knobs} lists both configurations).

Table~\ref{tab:compile} (Appendix~\ref{sec:app-full}) gives the \tool{}
stage time of both configurations on every benchmark of
Table~\ref{tab:speedups}; we summarize it here. On the 14 kernels,
\emph{Thorough}, the configuration behind our speedup table, takes a
median of 1.6\,s, and 10 of them compile in under 10\,s; the slowest are
Julia matmul and SHA-256, at 43\,s and 42\,s. \emph{Fast} takes a median of 0.36\,s and at
most 0.60\,s, against about 0.1\,s for a conventional \texttt{-O3}. On
the three repositories Thorough takes 116\,s (CryptoSwift), 77\,s
(zstd), and 347\,s (zlib), and Fast 63\,s, 17\,s, and 140\,s. These are massive repositories, and a developer often cares more about per-module latency as each module is compiled individually in the development loop. The median zlib module takes 4.2\,s
under Thorough and 1.1\,s under Fast, the median zstd module 1.3\,s
and 0.36\,s, and 25 of the 32 modules compile in under 10\,s under
Thorough, 30 under Fast. Zlib's \texttt{trees.c} is an outlier which takes
220\,s under Thorough and 103\,s under Fast; Figure~\ref{fig:modules}
(Appendix~\ref{sec:app-full}) shows the full per-module distribution.
The two dials trade that cost for proofs. \emph{Fast} keeps 382 of
Thorough's 885 proofs (\pct{43}) at two fifths of zlib's cost. 

\paragraph{Whole-compile cost.}
The latency that matters for a developer is the entire compile pipeline's time, not of
one optimizer run. A build runs a front end, the optimizer, and code
generation, and the front end is often a large share of it: Swift's, for
instance, runs its own optimizer before LLVM sees the code. Comparing
against \texttt{-O3} alone would therefore overstate the cost. We timed
every stage of the two pipelines the runtime experiments use, per module,
on the 13 benchmarks whose pipeline can take the pass output: the baseline (optimizing front end, \texttt{-O3}, code
generation) and the same pipeline with \tool{} sandwiched between two 
\texttt{-O3} runs. The median whole compile with \tool{} takes $1.49\times$ the
time of the baseline under Fast and $3.49\times$ under Thorough
(geometric means $1.83\times$ and $3.60\times$; ranges
$1.27$--$18.6\times$ and $1.50$--$44.4\times$). These are the numbers a
developer actually waits for, front end and code generation included,
and they are reasonable for an online pass: a typical build with Fast takes
little more than a conventional one, and even Thorough stays within a
small factor of it. 

\begin{figure}
\centering
\includegraphics[width=\columnwidth]{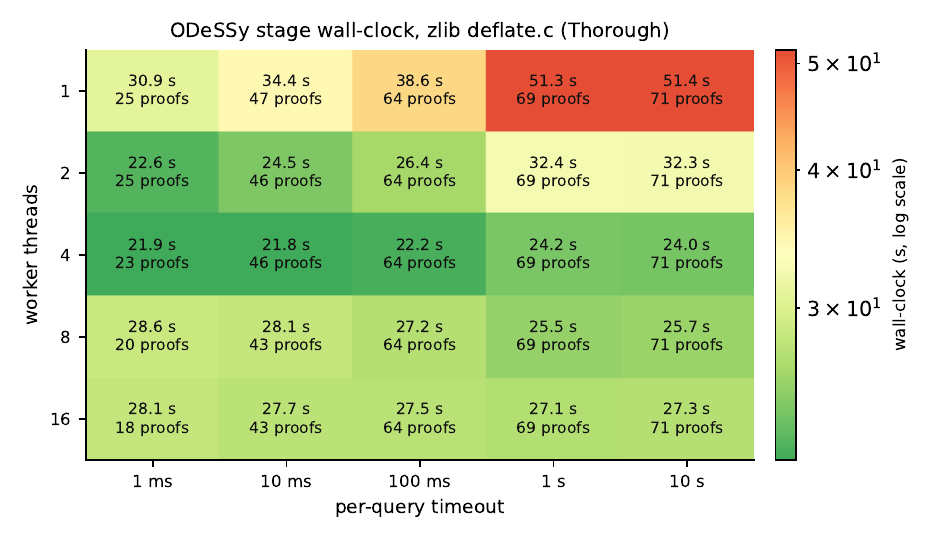}
\caption{\tool{}'s latency for zlib's \texttt{deflate.c} with a sweep over the number of threads and the solver timeout. The configuration is Thorough. Multi-threading and strict timeouts can make \tool{} up to $2.4\times$ faster. A strict timeout of 10\,ms keeps about two thirds of the proofs. Parallelism is the more effective measure for reducing latency without losing proofs: four workers are $2.1\times$ faster at the full 10\,s budget.}
\Description{A five-by-five heatmap, rows are 1, 2, 4, 8 and 16 worker
threads, columns are per-query timeouts from 1 millisecond to 10
seconds. Cells are colored from red for slow to green for fast on a log
scale and annotated with wall-clock seconds and proof counts. Wall-clock
falls from 51.4 seconds at one thread and 10 seconds to 21.8
seconds at four threads and 10 milliseconds; proof counts rise left to
right from 18 to 25 at 1 millisecond to 71 at 10 seconds.}
\label{fig:matrix}
\end{figure}

\paragraph{Parallelism and timeouts.}
Both dials make both configurations faster (Figure~\ref{fig:matrix}, zlib's
\texttt{deflate.c}). A strict per-query budget costs few proofs: \pct{93}
of them finish within 1\,s and \pct{75} within 100\,ms, because most queries
finish in milliseconds (Figure~\ref{fig:latency},
Appendix~\ref{sec:app-full}). Parallelism alone speeds \tool{} up by
$2.1\times$ at four workers with no proof lost (51.4\,s to 24.0\,s at
10\,s per query), and threads and budgets together by $2.4\times$ (to
21.8\,s at four workers and 10\,ms). It has a limit: discovery and fact
import, including the facts of inductive encoding, run in series, so
workers beyond four gain nothing.

\paragraph{Incremental compilation.}
A developer rarely compiles a project from scratch: most compiles follow
a small edit. The verdict cache (\S\ref{sec:cache}) makes these compiles
cheap, because \tool{} solves again only the functions that changed. To
measure it, we replayed the last 30 commits that change the sources of
zlib and of zstd, compiling each commit with the cache left by the previous
one. Without the cache, \tool{} under Thorough takes a median of 330\,s
per commit on zlib and 80\,s on zstd. With the cache, a commit that changes
checked code takes 23\,s and 6.1\,s ($14\times$ and $13\times$ faster), and
a commit that leaves the checked code unchanged takes 7.1\,s and 5.4\,s
($47\times$ and $15\times$ faster). The cache costs nothing measurable
when it is empty, and it does not change the result: every module of every
commit compiles to the same output with and without it. The one exception
is zlib's \texttt{deflate.c}, where the cache holds one extra proof,
recorded in a run where the query finished just under the 10\,s budget. So
a developer pays \tool{}'s full cost once, on the first compile of a
project, and afterwards only for the functions each edit changes.
Appendix~\ref{sec:app-cache} gives the per-module and per-commit results.

\paragraph{How to use \tool{}.}
Both configurations fit the developer's edit-and-compile loop. A build recompiles only
the module that changed, and a single module is cheap: under Thorough the
median module takes 4.2\,s in zlib and 1.3\,s in zstd, and the libraries'
main modules take under a minute. When a module or a whole code base is
too large for Thorough, Fast stays cheap: every kernel compiles in under
0.6\,s, and zstd and CryptoSwift in 17\,s and 63\,s. The developer can then
iterate with Fast and build the final binary with Thorough, the same
trade release builds already make with link-time and profile-guided
optimization. Fast is also useful on its own: where it keeps Thorough's
proofs, it keeps Thorough's speedups (Table~\ref{tab:fast},
Appendix~\ref{sec:app-full}).
\subsection{RQ4: How does super-analysis compare with superoptimization?}
\label{sec:results-souper}
For the scope of this study we compare one representative of each paradigm, \tool{} with Souper~\cite{sasnauskas2017souper}, the
synthesizing superoptimizer for LLVM IR, on the task this paper targets:
removing the safety checks of the two C libraries. We ran Souper in its
\emph{constant mode}, which only proves that a value is always the same
constant. The authors of Souper used this mode as an online optimizer, and compiled LLVM's own code base using it. In a smaller scale experiment,
we also ran Souper in its CEGIS synthesis mode.

\paragraph{Setup.} Souper pins LLVM~18, whose parser cannot read the IR
of newer front ends, so the comparison covers only the C libraries which can be compiled
with LLVM~18's clang. Each module is emitted by clang~18 and both tools read the emitted IR;
\tool{}'s runs used a packaged clang~18 whose output leaves exactly the same
checks as Souper's build (476 in zlib, 254 and 251 in zstd). \tool{}'s LLVM reads it as bitcode, which newer LLVM
versions are guaranteed to accept. Despite the shared frontend, each tool runs with its own pinned
LLVM optimizer and its own pinned Z3
(Souper: Z3 4.13; \tool{}: Z3 4.8.12), and is followed by its own LLVM's \texttt{-O3}. Souper's per-query budget is
10\,s, as for Thorough, with no external cache; its in-memory cache of
identical queries is left on, as shipped. Swift, Rust, and Julia are
excluded because their front ends emit IR newer than LLVM~18. Thorough and Fast use ten
threads.

\begin{figure}[t]
\centering
\includegraphics[width=0.9\linewidth]{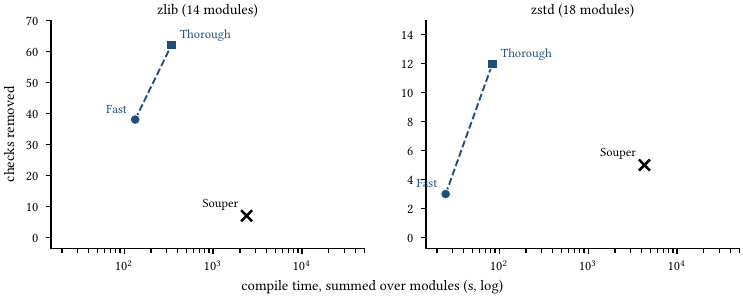}
\caption{Checks removed against compile time on every module of zlib and
zstd (x86, the same LLVM~18 IR for both tools). Time is summed over
modules and includes the \texttt{-O3} that follows each tool; checks are
counted against each tool's own \texttt{-O3} baseline. Thorough is faster and more effective than Souper in check removal.}
\Description{Two scatter panels, zlib and zstd, with compile time on a
logarithmic x-axis and checks removed on the y-axis. In each panel a dashed
line joins ODeSSy Fast (circle) and ODeSSy Thorough (square), and a black
cross marks Souper. zlib: Fast 133 seconds and 38 checks, Thorough 340
seconds and 62 checks, Souper 2382 seconds and 7 checks. zstd: Fast 25
seconds and 3 checks, Thorough 84 seconds and 12 checks, Souper 4296
seconds and 5 checks.}
\label{fig:souper-libs}
\end{figure}

\paragraph{Results.} Figure~\ref{fig:souper-libs} answers RQ4. On each
library Thorough beats Souper on both axes. On zlib,
Thorough removes 62 checks in 340\,s, where Souper removes 7 in
2{,}382\,s, and Fast removes 38 in 133\,s; most of Thorough's time is
spent on a single module, \texttt{trees.c} (Figure~\ref{fig:modules}).
On zstd, Thorough removes 12 checks in 84\,s, where Souper removes 5 in
4{,}296\,s, a $51.0\times$ difference in time, and Fast removes 3 in 25\,s. 

\begin{figure}[t]
\centering
\includegraphics[width=3.3in]{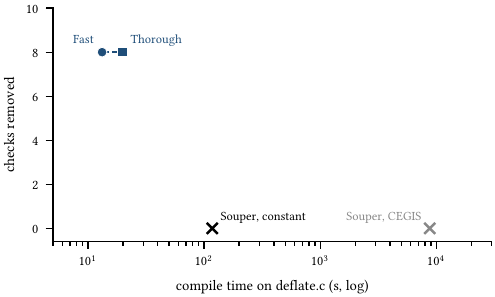}
\caption{zlib's \texttt{deflate.c}, the compressor: checks removed against
compile time, both tools on the same LLVM~18 IR (time includes the
\texttt{-O3} that follows each tool). Circle and square: \tool{}'s Fast
and Thorough; crosses: Souper in constant mode and in CEGIS synthesis
mode, neither of which removes a check.}
\Description{A scatter plot with compile time on a logarithmic x-axis and
checks removed on the y-axis. ODeSSy Fast removes 8 checks in 13 seconds
and ODeSSy Thorough 8 in 20 seconds, joined by a dashed line. Two crosses
sit at zero: Souper constant mode at 118 seconds and Souper CEGIS at 8771
seconds.}
\label{fig:souper-deflate}
\end{figure}

\paragraph{One module in detail.} We wanted to see how the more aggressive Souper mode, Counterexample-Guided Inductive Synthesis (CEGIS) performs. Figure~\ref{fig:souper-deflate} shows
the two tools on zlib's compressor, \texttt{deflate.c}, on the same
input. \tool{} removes 8 checks, in 13\,s under Fast and 20\,s under
Thorough. Souper's constant mode takes 118\,s and removes nothing. Its
synthesis mode ran for
2.4\,hours (8{,}771\,s) and also removed nothing; at that cost per
module we did not run it on the libraries.

Two cautions bound the comparison. It is one superoptimizer, on one task:
Souper is built to find general peephole improvements, and removing safety
checks is not its purpose, so the comparison measures the performance of the two paradigms on the problem this paper addresses only.
And we do not compare runtime: Souper removes checks in only three modules,
in each of which Thorough removes at least as many, and the two pipelines
end in different LLVM versions, so a runtime difference would mostly
measure LLVM~18 against LLVM~23 rather than the two tools.
\section{Related Work}
\label{sec:relwork}
\S\ref{sec:related} placed super-analysis among the paradigms of
aggressive optimization; here we compare it with work that shares its
tools or its target.

\paragraph{Combining analyses.}
Cousot and Cousot's reduced product~\cite{cousot1979systematic} combines abstract domains; Wegman and Zadeck~\cite{wegman1991constant} and Click and Cooper~\cite{click1995combining} combine specific analyses by hand; Lerner et al.~\cite{lerner2002composing} compose them automatically into a super-analysis; and Gulwani and Tiwari~\cite{gulwani2006combining} combine abstract interpreters through a logical product, in the tradition of the theory combination that underlies SMT solvers. \tool{} uses an SMT solver directly as the combiner, applied to the facts that existing compiler analyses already compute.

\paragraph{SMT for compiler analysis.}
Closer in spirit to our work is Taneja et al.~\cite{taneja2020knownbits},
who use an SMT solver to test LLVM's dataflow analyses. For each expression they compute the
\emph{most precise} answer an analysis could possibly give --- the
exact set of known bits, the tightest integer range --- and compare it
with LLVM's. LLVM's answers are sound but not always precise: their oracle finds a tighter integer range or
demanded-bits result in about \pct{8} of cases, and runs out of time on many of the rest. Their solver-based analysis
is slow --- 9\,s per expression for known bits, 16\,s for integer
ranges --- and they conclude it is ``too slow to use at compile time'', and is not viable as an online optimizer.
The cost comes from the question they ask. Finding the best abstract
value is a \emph{search} over the abstract domain: known bits takes
$2n$ solver queries for an $n$-bit value, and ranges take a binary
search, so one answer costs dozens of queries. \tool{} asks a different
and much cheaper question: one yes/no reachability query per check
site (decision rather than search), which is why our median query
takes milliseconds (\S\ref{sec:results-cost}).
Alive2~\cite{lopes2021alive2} applies a solver to compiler
\emph{correctness} rather than precision: it validates LLVM's
transformations by checking refinement between the IR before and after
each pass. Like \tool{}, it treats unencoded behavior as an
over-approximation, so a solver verdict is only trusted in one
direction.

\paragraph{Bounds-check elimination and sanitizer overhead.}
Removing redundant safety checks is a classic problem.
ABCD~\cite{bodik2000abcd} proves bounds checks redundant by solving
difference constraints over an extended SSA form,
HotSpot~\cite{wurthinger2007hotspot} eliminates them with range analysis
and loop hoisting, and LLVM's inductive range-check elimination splits a
loop around checks whose index is affine in the induction variable. Each
reasons in a single abstract domain chosen in advance; the proofs \tool{}
needs combine several domains at once (\S\ref{sec:conjunction}) and, for
heap-resident lengths, memory (\S\ref{sec:frame}). For sanitizers such as
AddressSanitizer~\cite{serebryany2012asan} and UBSan~\cite{ubsanDocs},
ASAP~\cite{wagner2015asap} removes the checks on the hottest code under a
user-chosen overhead budget, trading safety for speed, and
SanRazor~\cite{zhang2021sanrazor} removes checks it finds redundant with
other checks, using profiling as well as static analysis. \tool{} removes
only checks it proves can never fire, each backed by a solver proof.

\paragraph{Guards and loop versioning.}
Cloning a loop behind a runtime test is established practice: .NET's
RyuJIT clones loops to remove bounds checks, and LLVM's loop versioning
guards a copy by runtime alias checks. In both, the guard is taken from a
fixed syntactic pattern. Guard synthesis (\S\ref{sec:guards}) instead
obtains the guard from the solver, as an abductive
hypothesis~\cite{dillig2012abductive,dillig2013abductive}, and verifies
that it is sufficient and consistent before the loop is cloned.

\section{Conclusion}
\label{sec:conclusion}
We introduced solver-based super-analysis, an alternative paradigm of aggressive compiler optimization. Our instantiation of super-analysis uses a solver for analysis only, keeping transformations conventional. We show that this paradigm is effective: it can speed up our benchmarks up to $4.01\times$ on x86. We also show that it is relatively fast: super-analysis can be used online, and the cost of compilation can be controlled by parallelization and strict SMT timeout budgets. Our implementation, \tool{}, targets safety check removal and loop cloning under synthesized guards. Future work consists of extending this paradigm into other analyses and optimizations.
\section*{Data-Availability Statement}
The \tool{} pass, the benchmark sources and inputs, the harnesses that
build and time them, the scripts that produce every table and figure, and
the raw per-repetition measurements are available at
\url{https://github.com/AmiraliE1380/ODeSSy}, together with the pinned
toolchain versions and instructions to rebuild the environment. A Docker
image that rebuilds the pass and reruns the evaluation will be provided for
artifact evaluation.
\begin{acks}
We thank Zachary Tatlock for his feedback on this work and for suggesting
the verdict cache (\S\ref{sec:cache}). Experiments were run on CloudLab~\cite{cloudlab}. The authors used Claude
(Anthropic), an AI assistant, to help write parts of \tool{}'s code, the
experiment scripts, and the manuscript's text. The authors designed the
approach and the experiments, checked every result, and take full
responsibility for the paper's content.
\end{acks}
\bibliographystyle{ACM-Reference-Format}
\bibliography{odessy}
\clearpage
\appendix
\section{Reproducibility Summary}
Verdicts, output IR, and logs are invariant in \texttt{threads=N} by the
ticket-turnstile construction, up to queries that finish close to their time budget (\S\ref{sec:pipeline}); the acceptance test is a
\texttt{diff} of \texttt{threads=1} vs.\ \texttt{threads=8} outputs.
Hardware is specified to the CloudLab node type (\texttt{c220g2}), whose
full specification is published; all timed runs disable turbo and pin to
one socket; medians over shuffled interleaved repetitions are primary
throughout. 

\section{Additional Results}
\label{sec:app-full}
Measurement noise is small: the two baselines, which differ only by a
second \texttt{-O3} run, are a median of 0.10\% apart across the C,
Swift, and Rust benchmarks and at most 0.32\%, except on zstd's median run
(1.1\%) and zlib's two largest corpora (0.90\% at 64\,MB and 1.15\% at 256\,MB), where zlib's
$1.011\times$ is therefore within noise.
M-series measurements are in Appendix~\ref{sec:app-mseries}.

\begin{figure}[t]
\centering
\includegraphics[width=\textwidth]{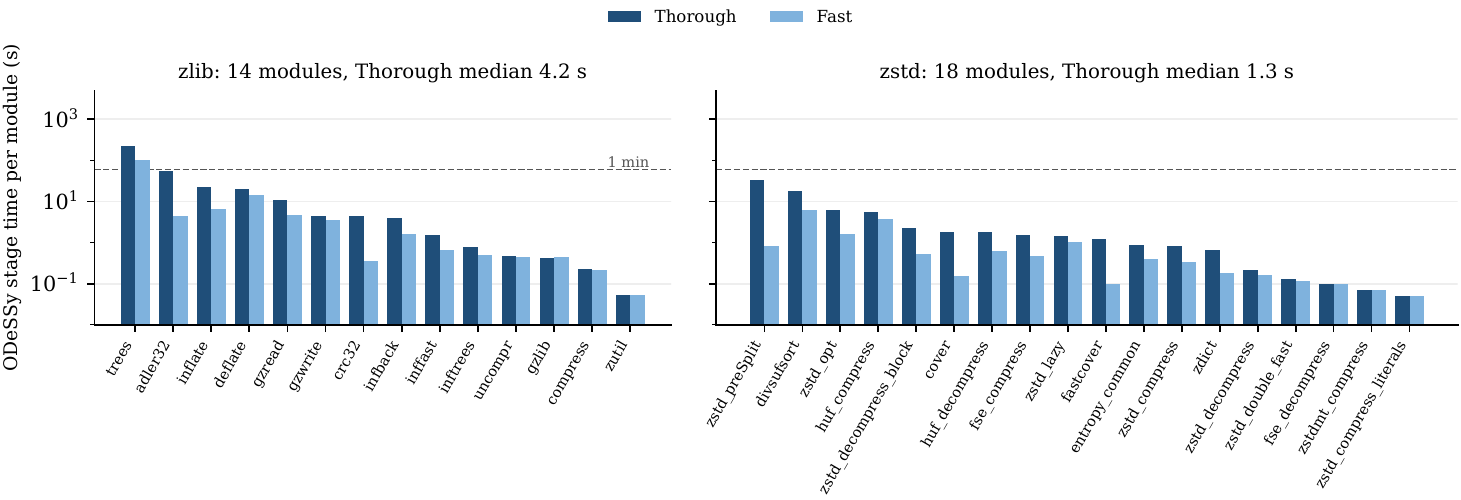}
\caption{Per-module \tool{} stage time for the two C libraries, modules
sorted by Thorough time, 10 threads for each tier. Under Thorough, 13 of zlib's 14
modules and all 18 of zstd's finish in about a minute or less (medians 4.2\,s
and 1.3\,s); the zlib total of Table~\ref{tab:compile} is set by one
outlier, \texttt{trees.c}.}
\Description{Two bar-chart panels on a logarithmic time axis from 0.01 to
over 1000 seconds. Left, zlib's 14 modules; right, zstd's 18 modules, each
sorted from slowest to fastest under Thorough. Each module has two bars
(Thorough, Fast). A dashed line marks
one minute. In zlib, trees rises far above the line under
Thorough; every other module sits near or below it.}
\label{fig:modules}
\end{figure}

Table~\ref{tab:fast} times the Fast configuration of
Table~\ref{tab:compile} on every C, Swift, and Rust benchmark where it
proves at least one check, with the same harnesses, workloads, and
repetitions as Table~\ref{tab:speedups}. Where Fast keeps the proofs that
matter it keeps Thorough's speedup (Rust matmul, adler32, zlib, md5, sha1);
where the speedup needs proofs that take longer than 10\,ms, Fast loses it
(sha256, zstd).
\begin{table}[t]
\caption{Fast against Thorough at runtime (x86 c220g2, medians of 30
shuffled repetitions, byte-identical output).
The Thorough column repeats Table~\ref{tab:speedups}. The two baselines
are within 0.3\% of each other, except on zstd (1.5\%) and zlib (1.1\%).}
\label{tab:fast}
\small
\begin{tabular}{llrr}
\toprule
Benchmark & Front end & Fast & Thorough\\
\midrule
matmul & Rust & $1.472\times$ & $1.453\times$\\
utf8 validator & Swift & $1.126\times$ & $1.025\times$\\
adler32 & Swift & $1.037\times$ & $1.039\times$\\
zstd, compression & C & $1.019\times$ & $1.057\times$\\
zlib & C & $1.009\times$ & $1.011\times$\\
sha256 & Swift & $1.001\times$ & $1.105\times$\\
md5 & Swift & $0.983\times$ & $0.983\times$\\
sha1 & Swift & $0.956\times$ & $0.956\times$\\
CryptoSwift & Swift & $0.952\times$ & $0.972\times$\\
lz77 & Swift & $0.694\times$ & $0.830\times$\\
\bottomrule
\end{tabular}
\end{table}
Table~\ref{tab:compile} gives the compile cost summarized in
\S\ref{sec:results-cost}, and Figure~\ref{fig:modules} its per-module
distribution on the two C libraries.
\begin{table}
\caption{Compile cost of two configurations on every benchmark of
Table~\ref{tab:speedups}. \emph{Thorough} is the
configuration used for the performance improvement tables (10\,s per query, ten
threads, inductive encoding and guard synthesis); \emph{Fast} keeps the
imported analyses, heap facts, and guard synthesis at 10\,ms per query
on 10 threads, without inductive encoding. For zlib and zstd the time reported in parentheses is the time taken by the largest module.}
\label{tab:compile}
\small
\setlength{\tabcolsep}{3pt}
\resizebox{\linewidth}{!}{%
\begin{tabular}{llrr|rr|rr}
\toprule
 & & & & \multicolumn{2}{c|}{Thorough} & \multicolumn{2}{c}{Fast}\\
Benchmark & Front end & Checks & \texttt{-O3} & Time & Proofs & Time & Proofs\\
\midrule
gemm, stdlib & Julia & 16 & 0.05 & 11.1 & 9 & 0.36 & 8\\
lz77 & Julia & 4 & 0.04 & 20.2 & 2 & 0.13 & 0\\
matmul & Julia & 3 & 0.02 & 42.7 & 3 & 0.17 & 0\\
sha256 & Julia & 16 & 0.06 & 42.2 & 8 & 0.55 & 0\\
lz77 & Rust & 4 & 0.04 & 1.8 & 2 & 0.13 & 0\\
matmul & Rust & 4 & 0.04 & 0.6 & 2 & 0.17 & 2\\
sha256 & Swift & 36 & 0.11 & 1.3 & 8 & 0.54 & 1\\
base64 & Swift & 27 & 0.09 & 4.8 & 7 & 0.57 & 0\\
lz77 & Swift & 27 & 0.10 & 5.2 & 7 & 0.36 & 3\\
crc32 & Swift & 37 & 0.11 & 1.4 & 11 & 0.60 & 0\\
adler32 & Swift & 38 & 0.10 & 1.3 & 2 & 0.56 & 1\\
sha1 & Swift & 24 & 0.12 & 0.7 & 2 & 0.36 & 2\\
md5 & Swift & 25 & 0.11 & 0.8 & 1 & 0.40 & 1\\
utf8 validator & Swift & 21 & 0.09 & 0.5 & 4 & 0.28 & 2\\
\midrule
CryptoSwift & Swift & 2696 & 4.75 & 116 & 378 & 63.1 & 150\\
zstd, 18 modules & C & 606 & 8.77 & 76.6 (33.8) & 151 & 16.8 (6.4) & 92\\
zlib, 14 modules & C & 1554 & 2.23 & 347 (220) & 288 & 140 (103) & 120\\
\midrule
Total & & 5138 & & & 885 & & 382\\
\bottomrule
\end{tabular}}
\end{table}
\begin{figure}[t]
\centering
\includegraphics[width=\textwidth]{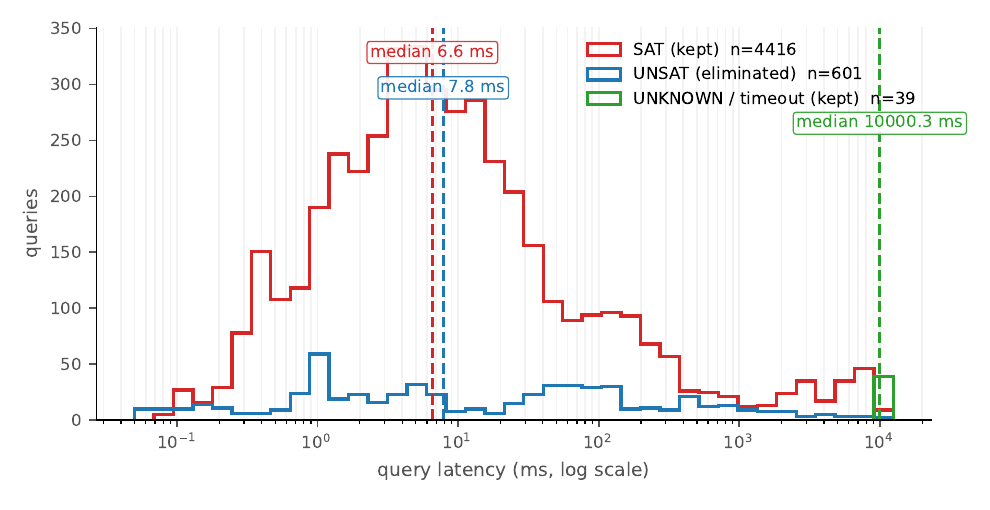}
\caption{SMT query latency by verdict class: every direct reachability
query (one per check) that the Thorough configuration poses on the 17
benchmarks of Table~\ref{tab:speedups} (5{,}056 queries; x86, ten
threads, 10\,s per query). The median proof takes 7.8\,ms, and \pct{75}
of proofs finish within 100\,ms and \pct{93} within 1\,s; refutations
are cheaper at the median (6.6\,ms). 39 queries reach the 10\,s budget
and keep their checks. }
\Description{A single histogram panel of SMT query latency on a
logarithmic x-axis in milliseconds, with query count on the y-axis. Three
overlaid outlines: SAT in red with 4416 queries, UNSAT in blue with 601
queries, and UNKNOWN in green with 39 queries at the 10 second cap.
Dashed vertical lines mark the medians at 6.6 milliseconds for SAT and
7.8 milliseconds for UNSAT. Both distributions spread from below 0.1
milliseconds to several seconds.}
\label{fig:latency}
\end{figure}
\begin{figure}
\centering
\includegraphics[width=0.92\textwidth]{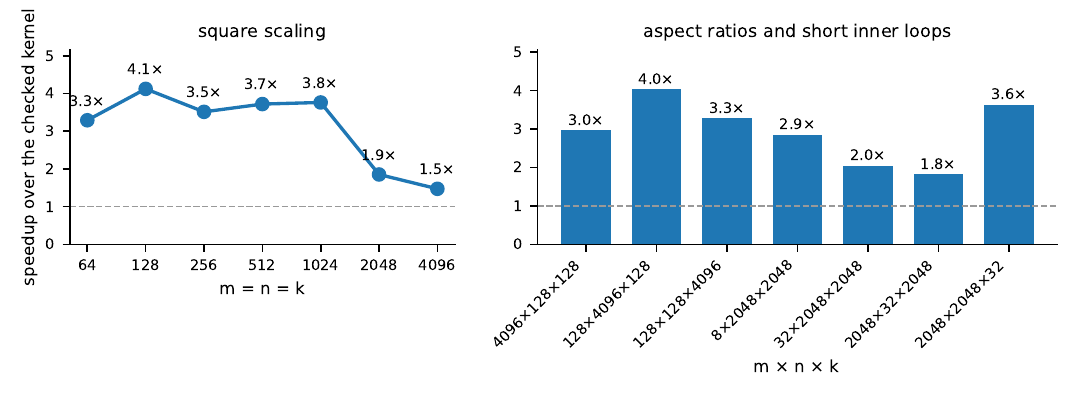}
\caption{GEMM speedup of \tool{}'s build (the two proven accesses
\texttt{@inbounds}, the other two checked) over the checked baseline, by
problem shape ($C = m{\times}n$, $A = m{\times}k$, $B = k{\times}n$;
medians of 11 repetitions, 5 at $4096^3$; bitwise-equal outputs at every
shape). \emph{Left:} square matrices. \emph{Right:} other aspect ratios.}
\Description{Two panels. Left, a line chart of speedup versus square
matrix dimension from 64 to 4096 on a log axis: 3.3x, 4.1x, 3.5x, 3.7x,
3.8x, 1.9x, and 1.5x. Right, a bar chart of seven non-square shapes with
speedups between 1.8x and 4.0x.}
\label{fig:gemm-sweep}
\end{figure}

\section{Definitional Encoding of Loop Bounds}
\label{sec:app-defs}
Two refinements make trip-count facts usable on vectorized loops, each
replacing a free variable with the definition of the code that computes
it. First,
trip-count bounds are stated in \emph{subtraction form},
\[
\den{\phi} - \mathit{start} \ \le_u\ s \cdot \mathrm{BTC}
\qquad\text{instead of}\qquad
\den{\phi} \ \le_u\ \mathit{start} + s \cdot \mathrm{BTC},
\]
where $\phi$ is a loop-header phi, $s$ its stride, and $\mathrm{BTC}$
the loop's back-edge-taken count. Modular subtraction cancels overflow
exactly, so the unit-stride bound needs no no-wrap side condition and
holds for the symbolic start values of remainder loops. Second, the
values these bounds mention (the vectorizer's resume indices and the code
that computes them) are encoded through the ordinary rules of
Table~\ref{tab:encoding} instead of as free variables. No ``trip count
$>0$'' is asserted anywhere: the countermodels that stand inside
zero-trip loops die because they violate these definitions. Each
addition holds in every execution, so soundness is unaffected. Neither
is specific to one benchmark; stdlib GEMM, whose loops the vectorizer
splits into vector, remainder, and runtime-checked versions, is where
they matter most: with the heap facts of \S\ref{sec:frame-machinery},
they take it from 0 to 9 of its 16 trap edges.

\section{Defined-Memory Precondition}
\label{sec:app-defined-memory}
LLVM lets a byte of memory hold poison or \texttt{undef} (a stored poison
value, or uninitialized memory), and a value loaded from such bytes may be
observed differently by each use: two \texttt{freeze}s of one poison load of
\texttt{\%A.size} may return different numbers, so a loop bound and a bounds
check derived from the same load are formally unrelated, and later LLVM
passes may compile the program accordingly. No intraprocedural analysis can
exclude this for memory a function receives from its caller. \tool{}
therefore separates two obligations. The caller is responsible for the
well-formedness of its inputs: loaded memory, in particular array metadata
such as lengths, holds defined values. Under this precondition, \tool{}
proves for each function that every check it removes cannot fire. The
precondition is the default; the parameter \texttt{poison-loads} drops it
and treats every loaded value as possibly poison, which makes \tool{}
faithful to LLVM's semantics without conditions.

The precondition fails only in programs that their own language already
rejects or leaves undefined. Two reads of one location can disagree in only
two ways: another thread writes between them, or the location holds no
defined value (it was never initialized, or a poison value was stored to
it), in which case LLVM may choose a different value at each use. The semantics of Julia, Swift, and Rust rule out the second: their runtimes always initialize array
metadata, and safe code cannot create poison. The first is a data race,
which safe Rust rejects at compile time and which Swift and Julia leave
undefined. For C, the precondition amounts to the absence of data races,
which are undefined behavior since C11, and of reads of uninitialized
memory; these are the properties that ThreadSanitizer and MemorySanitizer
check. The precondition is also not new: a manual \texttt{@inbounds}
annotation, such as those in Julia's stdlib GEMM, is sound only if the
array's length reads the same each time. Frontends could discharge the
precondition in the IR itself by marking such loads \texttt{!noundef},
which \tool{} already honors. GEMM, for example, keeps 9 of its 16 trap
edges removed and its full speedup under the precondition, and 3 without
it.

\section{Configuration Knobs}
\label{sec:app-knobs}
Every mechanism of \tool{} is a parameter of the pass,
\texttt{oracle-pass<\emph{knob};\emph{knob};\ldots>}, so that each
earlier configuration stays reproducible and each mechanism can be
ablated alone. Table~\ref{tab:knobs} lists them. With no knobs, the pass
runs \emph{Thorough} = \texttt{heavy;frame;ind;mv;narrow;timeout=10000},
on one worker per hardware thread; the presets \texttt{thorough} and
\texttt{fast} name the two configurations of Table~\ref{tab:compile}
(\emph{Fast} = \texttt{heavy;frame;mv;narrow;timeout=10}), and the
\texttt{no-} knobs remove one mechanism from either. Naming any mechanism
knob explicitly selects exactly the mechanisms named, so every
configuration in this paper is reproducible from its knob string; our
experiments set \texttt{threads=} explicitly (10 for compile cost, 1 for
the runtime builds).
\begin{table}[h]
\caption{The knobs of \tool{}. Defaults in brackets: with no knobs the
pass runs Thorough.}
\label{tab:knobs}
\small
\resizebox{\linewidth}{!}{%
\begin{tabular}{lp{0.72\linewidth}}
\toprule
Knob & Effect\\
\midrule
\multicolumn{2}{l}{\emph{Presets}}\\
\texttt{thorough} & [default] \texttt{heavy;frame;ind;mv;narrow}, 10\,s per query (\S\ref{sec:results-cost}).\\
\texttt{fast} & \texttt{heavy;frame;mv;narrow}, 10\,ms per query (\S\ref{sec:results-cost}).\\
\texttt{no-ind}, \texttt{no-mv}, \texttt{no-frame}, \texttt{no-narrow} & Remove one mechanism from the default or a preset.\\
\midrule
\multicolumn{2}{l}{\emph{Fact sources}}\\
\texttt{light} / \texttt{heavy} & [\texttt{heavy}] Light encodes the context alone;
heavy also imports LLVM's analyses as facts (vacuity, SCEVSYM,
and the header-phi rules), \S\ref{sec:system-facts}.\\
\texttt{frame} & [on] Heap-invariance facts (\S\ref{sec:frame-machinery}) and the definitional encoding of Appendix~\ref{sec:app-defs}.\\
\texttt{ldeq} & [off] Load equality within a basic block (\S\ref{sec:system-facts}).\\
\texttt{nophiinv} & [off] Disables the header-phi rules (for ablation).\\
\texttt{poison-loads} & [off] Drops the assumption that loaded values are not poison (\S\ref{sec:encoding-impl}).\\
\texttt{ind} & [on] Inductive body encoding (\S\ref{sec:app-machinery}).\\
\midrule
\multicolumn{2}{l}{\emph{Guard synthesis} (\S\ref{sec:guards-impl})}\\
\texttt{mv} & [on] Guard synthesis and loop cloning with templates T1--T3.\\
\texttt{mv-light} & [off] The same with templates T1 and T2 only.\\
\texttt{mv-sane=}$k$ & [62] Largest rung $2^k$ of template T2's sane-range ladder ($8 \le k \le 62$).\\
\texttt{narrow}[\texttt{=}$b$] & [on; $b$=16] Certifies guards whose queries multiply 64-bit
values by re-encoding them at 32 bits with overflow flags, sound only under the guard.\\
\midrule
\multicolumn{2}{l}{\emph{Cost and audit}}\\
\texttt{timeout=}$t$ & [10000] Per-query solver budget in milliseconds.\\
\texttt{threads=}$n$ & [0] Worker threads; 0 means one per hardware thread. Verdicts do not depend on $n$
unless a query finishes close to its budget, where contention decides whether it completes (\S\ref{sec:pipeline}).\\
\texttt{vacuity} & [off] Re-solves every \textsc{unsat} without the trap condition and refuses
the proof if the context alone is contradictory (\S\ref{sec:audit}).\\
\texttt{traps=}$a$\texttt{:}$b$ & [none] Further callee names treated as checks: \texttt{panic}
for Rust, \texttt{bounds\_error:boundserror} for Julia (\S\ref{sec:pipeline}).\\
\texttt{profile}[\texttt{=}$t$] & [off; $t$=1000] Dumps every query slower than $t$\,ms as SMT-LIB.\\
\midrule
\multicolumn{2}{l}{\emph{Experiment instruments} (not optimizations)}\\
\texttt{mv-keep} & Clones and guards as \texttt{mv} does but leaves every check in the guarded
copy (pipeline (ii) of Table~\ref{tab:ablation}).\\
\texttt{mv-nofold} & As \texttt{mv-keep}, with each kept check given its own trap so its fate
can be traced.\\
\texttt{tag-checks} & Gives every check its own identity and stops, so pipelines can be
compared check by check.\\
\bottomrule
\end{tabular}}
\end{table}

\section{Verdict Cache Results}
\label{sec:app-cache}
We measure the verdict cache (\S\ref{sec:cache}) in two experiments,
both in the Thorough configuration with 10 threads on the \texttt{c220g2}
node: a replay of each library's recent history, and an identical rebuild
of every benchmark.

\paragraph{Commit replay.} For zlib and zstd we replayed the last 30
first-parent commits that change the library's sources, ending at the
benchmark versions. Each commit is compiled twice: once with the cache left
by the previous commit (the first commit only fills it), and once without a
cache. Table~\ref{tab:cache-replay} reports \tool{}'s time per commit, for
each library and for its most expensive modules. Times are medians over the
replayed commits, and ratios are speedups over the compile without a cache.
The \emph{changed} column uses the commits that changed that row's checked
functions, and the \emph{identical} column the commits that did not; a dash
means no replayed commit changed that module.

\begin{table}[h]
\centering\small
\caption{\tool{} time per commit with and without the verdict cache.}
\label{tab:cache-replay}
\begin{tabular}{lrrr}
\toprule
Benchmark & No cache & Changed, warm cache & Identical, warm cache\\
\midrule
zlib (whole library) & 330\,s & 23.0\,s ($14\times$) & 7.06\,s ($47\times$)\\
\quad \texttt{trees} & 206\,s & --- & 3.14\,s ($66\times$)\\
\quad \texttt{adler32} & 55.3\,s & --- & 0.25\,s ($223\times$)\\
\quad \texttt{inflate} & 21.2\,s & 21.4\,s ($1.0\times$) & 0.69\,s ($31\times$)\\
\quad \texttt{deflate} & 19.6\,s & 17.1\,s ($1.1\times$) & 1.13\,s ($17\times$)\\
zstd (whole library) & 80.1\,s & 6.14\,s ($13\times$) & 5.40\,s ($15\times$)\\
\quad \texttt{zstd\_preSplit} & 32.3\,s & --- & 0.07\,s ($455\times$)\\
\quad \texttt{divsufsort} & 22.7\,s & 18.0\,s ($1.3\times$) & 1.14\,s ($20\times$)\\
\bottomrule
\end{tabular}
\end{table}

Most commits leave most checked functions untouched, so the cache serves
almost everything: over the replayed commits, 96.5\,\% of zlib's and
98.5\,\% of zstd's checked functions came from the cache, and 22 and 21 of
the 29 commits after the first needed no solving at all. A commit that
leaves a library's checked functions identical compiles $47\times$ (zlib)
and $15\times$ (zstd) faster, and one that changes them still compiles
$14\times$ and $13\times$ faster, because only the changed functions are
solved again. The module rows of Table~\ref{tab:cache-replay} show both
ends: \texttt{trees}, \texttt{adler32}, and \texttt{zstd\_preSplit}, the
most expensive modules, were never changed and drop to at most a few
seconds, while a commit that changes \texttt{inflate},
\texttt{deflate}, or \texttt{divsufsort} gains little in that module
($1.0$--$1.3\times$), since the changed functions are the costly ones.

Every module's output with the cache equals its output without it in every
commit, with one exception: from the commit that changed five of
\texttt{deflate.c}'s functions on, its cached output removes one more check.
In that commit's cached run only the five changed functions were solved,
and one of their queries proved its check in 9.9\,s; in the uncached
compile, where all 19 functions compete for the processor, the same query
reaches the 10\,s budget. Both outputs are proved; the cache replays the
proof it recorded.

\paragraph{Identical rebuild.} We also compiled each of the 67 inputs of
our benchmarks three times: without the cache, with an empty cache (cold),
and again with the cache that run left (warm). Summed over the inputs,
\tool{} takes 871\,s without the cache, 852\,s cold, and 251\,s warm, so
keying and recording cost nothing measurable. The warm compile's IR equals
the cold compile's and the uncached compile's on 65 of the 67 inputs; the
other two are Julia SHA-256, which the cache does not store and whose
queries finish near the budget. Functions whose guard has a symbolic bound
(template T3) are not stored and are solved again on every compile; this is
why Julia matmul and SHA-256 do not benefit.

\section{M-series Results}
\label{sec:app-mseries}
All runs on one Apple M-series laptop (no pinning; medians of shuffled
interleaved repetitions; byte-identical outputs required), with the
same front ends as on x86: the pinned swift.org Swift 6.3.3 toolchain,
Julia 1.12.6, and Rust 1.97.1. Every row uses the
configuration of Table~\ref{tab:speedups} (Thorough at one thread,
10\,s per query) and measures its ceiling in the same run.

\begin{table}[h]
\caption{M-series runtime results: speedup of \tool{}'s build over the
checked build, and the ceiling (checked over unchecked) from the same
run. Rows and footnotes follow Table~\ref{tab:speedups}.}
\label{tab:app-runtime}
\small
\begin{tabular}{llrr}
\toprule
Benchmark & Front end & Speedup & Ceiling\\
\midrule
gemm, $n{=}512^{a}$ & Julia & $4.143\times$ & $4.189\times$\\
lz77$^{a,b}$ & Julia & $2.595\times$ & $2.597\times$\\
lz77$^{a}$ & Julia & $1.480\times$ & $2.597\times$\\
lz77 & Swift & $1.318\times$ & $1.292\times$\\
matmul, $n{=}128^{a}$ & Julia & $1.232\times$ & $1.240\times$\\
zstd, compression & C & $1.146\times$ & $1.145\times$\\
base64 & Swift & $1.141\times$ & $1.136\times$\\
sha256 & Swift & $1.087\times$ & $1.007\times$\\
md5 & Swift & $1.021\times$ & $0.968\times$\\
matmul, $n{=}512^{a}$ & Julia & $1.020\times$ & $1.021\times$\\
sha1 & Swift & $1.015\times$ & $0.977\times$\\
CryptoSwift & Swift & $1.012\times$ & $1.003\times$\\
crc32 & Swift & $1.006\times$ & $1.001\times$\\
matmul & Rust & $1.003\times$ & $0.977\times$\\
zstd, decompression & C & $1.000\times$ & $0.999\times$\\
utf8 validator & Swift & $0.991\times$ & $0.991\times$\\
zlib & C & $0.990\times$ & $0.999\times$\\
adler32 & Swift & $0.983\times$ & $1.055\times$\\
sha256$^{a}$ & Julia & $0.922\times$ & $0.923\times$\\
lz77 & Rust & $0.745\times$ & $0.965\times$\\
\bottomrule
\end{tabular}\\
\raggedright\footnotesize
$^{a}$Julia rows time a separate copy of the unchanged kernel, transformed
by hand exactly as \tool{}'s proofs license (as in
Table~\ref{tab:speedups}).
$^{b}$Guard synthesized under a 60\,s solver budget; at the default
10\,s budget, one of the two loads' checks is proven (Julia lz77 below).
Medians of 30 repetitions (21 for Julia).
zstd: 512\,MB corpus. zlib: 256\,MB input, 20 runs (8 and 64\,MB:
$0.995\times$ and $0.992\times$).\par
\end{table}

\end{document}